\documentclass[useAMS,usenatbib]{mn2e}
\usepackage{amsmath,fleqn,graphicx,amssymb}
\usepackage{multirow}
\renewcommand{\[}{\begin{equation}}
\renewcommand{\]}{\end{equation}}
\def\p{\partial}\def\i{{\rm i}}

\let\boldgrk=\gkvecten
\let\boldgrksc=\gkvecseven

\def\gkthing#1{{\mathchoice%
	{\hbox{{\boldgrk\char#1}}}
	{\hbox{{\boldgrk\char#1}}}
	{\hbox{{\boldgrksc\char#1}}}
	{\hbox{{\boldgrksc\char#1}}}}}

\def\vtheta{\gkthing{18}}

{\newif\ifnotend
\notendtrue
\def\veclist{ABCDEFGHIJKLMNOPQRSTUVWXYZabcdefghijklmnopqrstuvwxyz.}
\def\top#1#2.{#1}
\def\tail#1#2.{#2.}
\loop\expandafter\xdef\csname v\expandafter\top\veclist\endcsname%
{{\noexpand\bf\expandafter\top\veclist}}
\edef\veclist{\expandafter\tail\veclist}
\if\veclist.\notendfalse\fi\ifnotend\repeat}
\def\d{{\rm d}}

\def\cJ{{\cal J}}

\def\bolOm{\mbox{\boldmath$\Omega$}}
\def\vOmega{\bolOm}

\def\Gyr{\,\mathrm{Gyr}}
\def\Myr{\,\mathrm{Myr}}
\def\kpc{\,\mathrm{kpc}}

\def\kms{\,\mathrm{km\,s}^{-1}}

\def\msun{\,{\rm M}_\odot}

\def\e{\mathrm{e}}

\def\fracj#1#2{{\textstyle{#1\over#2}}}

\def\TM{{\sc tm}}

\def\Omegap{\Omega_{\rm p}}

\title[Angle-action variables for orbits trapped at a Lindblad resonance]
{Angle-action variables for orbits trapped at a Lindblad resonance}

\author[James Binney]{
  James Binney$^1$\thanks{E-mail: binney@physics.ox.ac.uk}\\  
  $^1$Rudolf Peierls Centre for Theoretical Physics, Clarendon Laboratory,
  Parks Road,
  Oxford, OX1 3PU, UK
}

\begin{document}
\maketitle

\begin{abstract}
The conventional approach to orbit trapping at Lindblad resonances via a pendulum
equation fails when the parent of the trapped  orbits is too circular. The
problem is explained and resolved in the context of the Torus Mapper and a
realistic Galaxy model. Tori are computed for orbits trapped at both the
inner and outer  Lindblad
resonances of our Galaxy. At the outer Lindblad resonance, orbits are
quasiperiodic and can be accurately fitted by torus mapping. At the inner
Lindblad resonance, orbits are significantly chaotic although far from
ergodic,
and each orbit explores a small range of tori obtained by torus mapping.
\end{abstract}


\section{Introduction}

It has long been recognised that angle-action variables, $(\vtheta,\vJ)$ are
valuable tools for galactic dynamics
\citep[e.g.][]{LLKa72,Kalnajs1977,Weinberg2001}.  Over the last decade, the
use of angle-action variables has become more widespread on account of the
arrival of algorithms for computing the transformation between these rather
abstract variables and ordinary phase-space variables $(\vx,\vv)$. All these
algorithms derive from our ability to solve the Hamilton-Jacobi equation for
the St\"ackel potentials \citep{deZ85} or limiting cases of them, but
a variety of stratagems make it possible to obtain good approximations to
angle-action coordinates for most plausible galactic potentials. One
stratagem is the St\"ckel Fudge \citep{JJB12:Stackel,SaJJB15:Triaxial},
another is torus mapping \citep[][and references therein]{JJBPJM16}, and a third is construction of the
generating function by orbit integration \citep{SaJJB14}. These techniques
are all limited to orbits that are qualitatively the same as orbits supported
by St\"ackel potentials.

The phenomenon of resonant trapping gives rise to orbits in real potentials
that are qualitatively different from any orbit in a St\"ackel potential, and
we have reason to believe that such orbits are astronomically important
\citep[e.g.][]{WD00:OLR,MonariHerc2017,Perez2017}.
\citealt[(hereafter B18)]{Binney2018} gave algorithms for the construction of
angle-action variables  for orbits that are trapped
at the principal resonances of a realistic model of the Galaxy's bar. The
paper focused of trapping at the corotation and outer Lindblad resonances
(OLRs),
but its approach is more widely applicable. Indeed, the code released with
the paper was directly applicable to orbits trapped at the inner Lindblad
resonance (ILR). C++ code that computes tori trapped at the principal
resonances of a disc was released in the form of an upgrade to the Torus
Mapper (hereafter \TM) \citep{JJBPJM16}.

In the course of a study of signatures of resonant trapping in velocity space
at the Sun \citep{Binney:Vspace}, it emerged that the standard approach
adopted by B18 fails for certain nearly circular orbits. The principal aim of
this paper is to explain what this problem is, and to show how it can be
solved.  The performance of the version of \TM\ that resolves this problem is
illustrated by tori that straddle the standard and new regimes at both the
ILR and OLR.

Paper is organised as follows. Section~\ref{sec:Phi} specifies the model of
the Galactic disc and bar that is employed. Section~\ref{sec:slow} explains
why the standard pendulum equation relied on by B18 breaks down and presents
a more robust algorithm.  Section~\ref{sec:OLR} applies the new algorithm to
orbits trapped at OLR, while Section~\ref{sec:ILR} applies it to orbits
trapped at the ILR.  Section~\ref{sec:Etheory} connects the angle-action
approach to elementary theory of perturbed orbits, and
Section~\ref{sec:conclude} sums up.

\section{The potential and units}\label{sec:Phi}

We need a reasonable model of our Galaxy's potential $\Phi(\vx)$. As in
B18 we assemble $\Phi$  by adding a quadrupole to an axisymmetric part
\[
\Phi(R,z,\phi)=\Phi_0(R,z)-\Phi_2(R,z)\cos(2\phi),
\]
 where the minus sign ensures that the bar's long axis is $\phi=0$. It is
natural to adopt for $\Phi_0$ a potential that has been fitted to a variety
of observational data under the assumption that the Galaxy comprises thin and
thick discs, and axisymmetric bulge and dark halo. As in B18 we for
the most part adopt the ``best'' potential of \cite{PJM11:mass}.

\begin{figure}
\centerline{\includegraphics[width=.8\hsize]{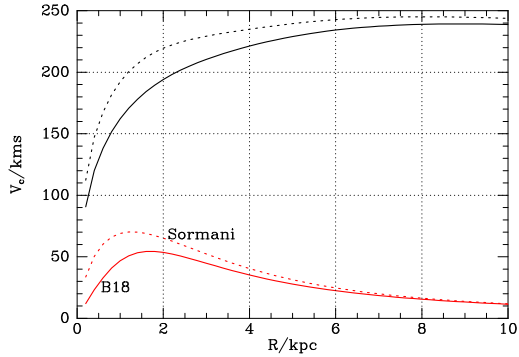}}
\caption{Full black curve: circular-speed of the ``best'' potential
of McMillan (2011). Dotted black curve: circular-speed of an axisymmetric
potential that is consistent with the existence of $x_2$ orbits in
combination with the S15 bar. Full red
curve: $(\p\Phi_2/\p\phi)^{1/2}$ from the B18 bar model. Dashed red curve
$(\p\Phi_2/\p\phi)^{1/2}$  from the S15 bar model.}\label{fig:quads}
\end{figure}

\begin{figure}
\centerline{\includegraphics[width=.8\hsize]{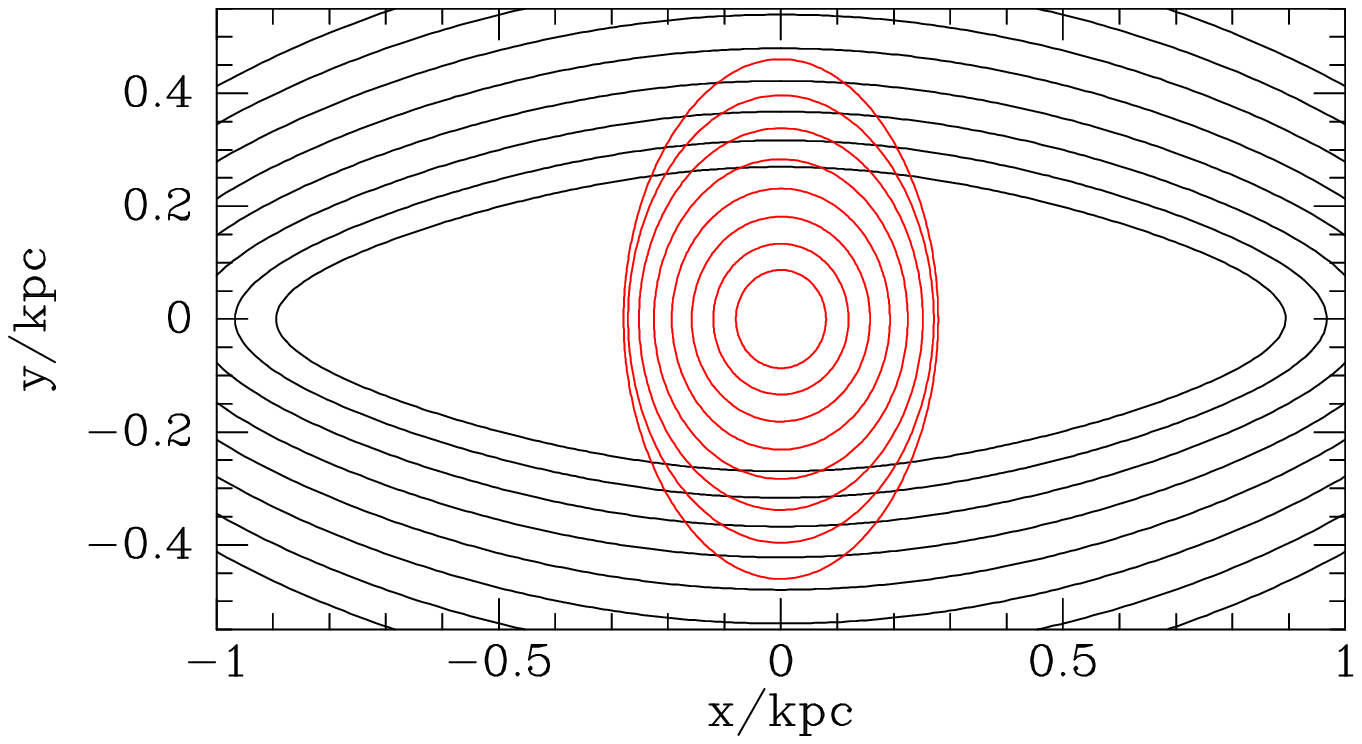}}
\centerline{\includegraphics[width=.8\hsize]{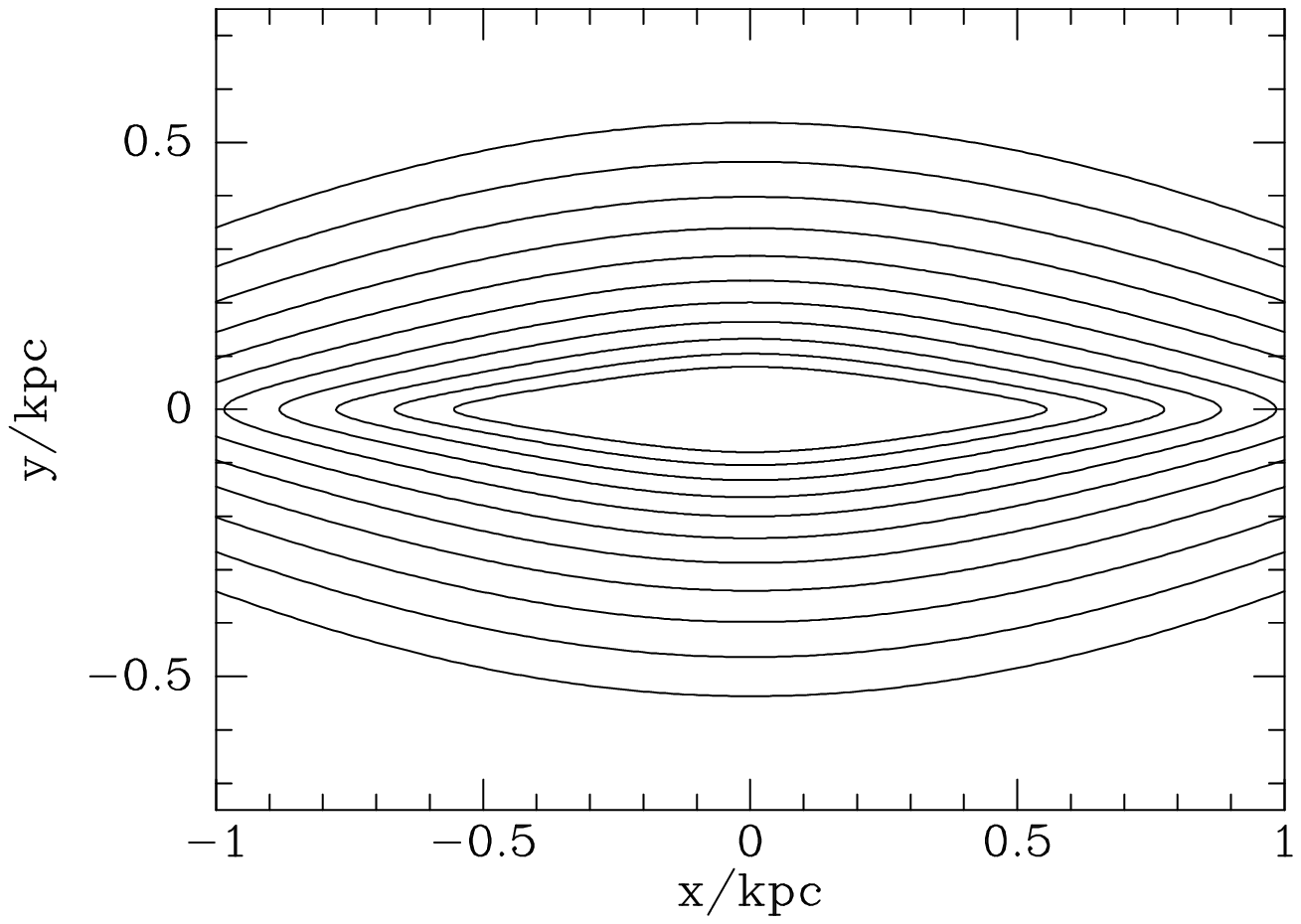}}
\caption{Closed orbits in the B18 and S15 potentials.}\label{fig:BSclosed}
\end{figure}

\begin{figure}
\centerline{\includegraphics[width=.8\hsize]{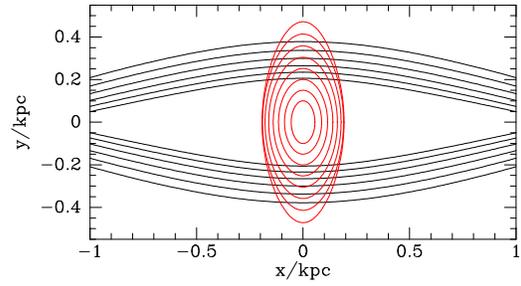}}
\caption{Closed orbits in the S15 bar combined with a more massive bulge. The
black dashed curve in Fig.~\ref{fig:BSclosed} shows the circular speed of the
axisymmetric component.}\label{fig:Sclosed}
\end{figure}

\citealt[][(hereafter S15)]{SormaniIII} constrained the bar's contribution to $\Phi$ with
hydrodynamical simulations of the flow of gas in the Galactic plane; these
simulations yielded longitude-velocity plots that could be compared with
observations in the $21\,$cm and $2.6\,$mm lines of hydrogen and CO.
S15 adopted a simple functional form for $\rho$ and obtained
$\Phi_2$ by integrating Poisson's equation.  B18 did not use the resulting
potential directly but instead adopted a simple functional form for $\Phi_2$,
derived the corresponding density from $\rho=(4\pi G)^{-1}\nabla^2\Phi$ and chose
the strength of the quadrupole component so that at large $R$ it matched that
of S15. This procedure guarantees good agreement between the B18
and S15 potentials at the OLR, but allows significant disparity between the
potentials at the ILR. Since we here present results at the ILR as well as
the OLR, it is important to compare the non-axisymmetric forces implied by
each bar model as a function of radius. Fig.~\ref{fig:quads} does that by
showing the circular-speed curve from the axisymmetric model alongside the
azimuthal analogues, $(\p\Phi_2/\p\phi)^{1/2}$, from the two bar models. These
agree by construction at large $R$, but
at small $R$ the S15 curve peaks sooner and higher.

The bottom two panels of Fig.~\ref{fig:BSclosed} shows the closed orbits in
these two model potentials. The greater quadrupole strength of the S15
potential at small radii makes the $x_1$ orbits (plotted in black) more
elongated than in the B18 model, and completely eliminates the $x_2$ orbits
(plotted in red). A viable bar model requires $x_2$ orbits because dense gas
moving on these orbits forms the heart of the Central Molecular Zone
\citep[e.g.][]{Ferriere2007}.
Fig.~\ref{fig:Sclosed} shows closed orbits when the S15 bar model is combined
with an axisymmetric model that has a more massive bulge and therefore a more
steeply rising circular speed curve. The full black curve in the top panel of
Fig.~\ref{fig:quads} shows the circular speed curve of the
\cite{PJM11:mass} potential that does not produce $x_2$ orbits when combined
with the S15 bar, while the black dashed curve shows a more steeply rising
circular speed curve obtained by increasing the central density of the bulge
from $9.56$ to $15.6\times10^{10}\msun\kpc^{-3}$.
Fig.~\ref{fig:Sclosed}
shows that with the more massive bulge the S15 bar permits $x_2$ orbits.

\subsection{Units}

In \TM\ the natural unit of distance is kpc and the natural unit of velocity
is $\kpc\Myr^{-1}=978\kms$, with the consequence that the natural units of
frequency and action are $\Myr^{-1}$ and $\kpc^2\Myr^{-1}$, respectively. In the
following units will not be given on the understanding that quantities are
expressed in \TM's natural units.

\section{Motion in the slow plane}\label{sec:slow}

By torus mapping one can foliate phase space by 3-tori on which a given
axisymmetric 
Hamiltonian is nearly constant. In fact, a foliation by tori {\it defines} a
Hamiltonian 
\[\label{eq:defsH0}
\overline{H}(\vJ)\equiv{1\over(2\pi)^3}\int\d^3\vtheta\,H(\vtheta,\vJ)
\]
for which the foliation's tori are orbital tori. In the following $\vOmega$
denotes the vector formed by the frequencies $\p\overline{H}/\p J_i$ that
this integrable Hamiltonian defines. Motion in the full Hamiltonian can then be
studied as the effect of the perturbation 
\[\label{eq:defsh1}
H_1(\vtheta,\vJ)\equiv H(\vtheta,\vJ)-\overline{H}(\vJ)
\]
on the integrable motion in $\overline{H}$ for which we have angle-action
coordinates. 

We express $H_1$ as the Fourier series 
\[\label{eq:Fh1}
H_1(\vtheta,\vJ)=\sum_{\vn}H_\vn(\vJ)\e^{\i\vn\cdot\vtheta},
\]
where the $\vn$ are 3-vectors with integer components.  Even a small
amplitude $H_\vn$ can qualitatively change the dynamics if it acts in the
same sense for a long time, as it will do if the corresponding frequency
$\vn\cdot\vOmega$ is small. One finds that remarkably accurate results can be
obtained by retaining in the sum of equation (\ref{eq:Fh1}) only terms in
which $\vn$ is equal to some vector $\vN$ and its integer
multiples.\footnote{By the reality of $H_1$ we must always include also
$\vn=-\vN$ and \cite{Binney2016} included also $\vn=\pm2\vN$.} B18 studied
the cases $\vN=(0,0,1)$ (corotation) and $\vN=(1,0,2)$ (OLR).
The complex number $H_\vn$ was reduced to its modulus and phase by writing
\[
H_\vn=h_\vn\e^{\i\psi_\vn}.
\]

Dropping all terms except those involving $\vN$ essentially reduces the motion
to that in the plane spanned by the `slow' angle and action 
\[
\theta_1'\equiv\vN\cdot\vtheta\hbox{ and } J_1'\equiv {J_1\over N_1}.
\] 
$J_1'$ is not a constant of motion and its conjugate variable
$\theta_1'$ does not increase linearly with time. Linear combinations of the
old actions remain approximate constants of motion, however:
\[
J_3'=J_3-{N_3\over N_1}J_1
\]
and $J_2'$ which is obtained by replacing 3 with 2 in this  formula.

A third real number in addition to $h_\vN$ and $\psi_\vN$ plays an important
role in the dynamics, namely
\[
G\equiv{\p\Omega_1'\over\p J_1'}={\p^2\overline{H}\over\p J_1^{\prime2}}
\]
evaluated when $\vN\cdot\vOmega=0$.

\begin{figure}
\centerline{\includegraphics[width=.9\hsize]{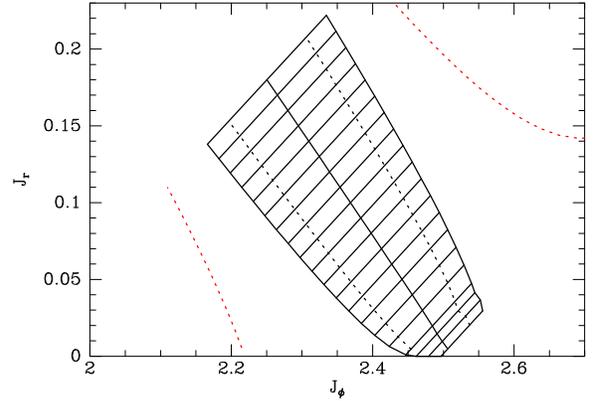}}
\caption{Orbits trapped at the OLR in action space. The centre line of the ladder marks
the  actions of orbits that satisfy the OLR resonant condition
$\Omega_r+2(\Omegap-\Omega_\phi)$ in the axisymmetric potential. The ladder
is bounded by the extremes of the unperturbed actions that are reached by
orbits trapped at OLR. The dashed lines within the ladder show the actions of
orbits that just circulate inside or outside OLR. The red dashed lines show
the predictions of the classical pendulum equation for the extremes in
actions achieved by trapped orbits.}\label{fig:OLR_ladder}
\end{figure}

\subsection{The problem}

The critical weakness of the standard algorithm used by B18 can be anticipated
from Fig.~\ref{fig:OLR_ladder}, which is a plot of the $(J_\phi,J_r)$ plane in
the vicinity of the OLR. A ladder sloping from lower right to upper left is
evident. Its centre line marks the axisymmetric
tori that are in perfect resonance with the bar because on them
$2(\Omegap-\Omega_\phi)=\Omega_r$. The ladder's `rungs' are formed by
the lines
\[
J_3'\equiv J_\phi-2J_r=\hbox{constant}
\]
along which libration carries an orbit in the presence of a bar: the bar
changes both a star's angular momentum and its radial action, but in such a
way that $J_3'$ is (approximately) invariant at the value associated with its
rung. The problem we have to address is posed by rungs that cross the
ladder's central line at a small value of $J_r$: the standard pendulum
equation used by B18 implies that such rungs carry the star to negative
$J_r$, which is not physically possible.

The underlying mathematical problem is this. Orbit-averaging of the
Hamiltonian $H$ essentially reduces the dynamics to motion in the slow plane.
In the case of the OLR, the natural coordinates for this plane are $J_1'=J_r$
and $\theta_1'=\theta_r+2\theta_\phi$. The standard argument is that the difference
$\Delta\equiv J_1'-J_{1\,\rm res}'$ between the current slow action and that
of the perfectly resonant torus satisfies a modified pendulum equation, and
if $J_{1\,\rm res}'$ is sufficiently small, the amplitude of the pendulum's
oscillations can be large enough to make $J_{1\,\rm res}'+\Delta<0$. Given
that this problem arises in the regime of nearly circular orbits, in which
the classical epicycle approximation should apply, it cannot define a
fundamental problem with the use of linear theory, but must arise from a
removable cause.

\subsection{The solution}

The problem is the coordinate singularity at the origin of the
$(J_1',\theta_1')$ phase plane. At sufficiently large $J_{1\,\rm res}'$, the
motion in $(\theta_1',J_1')$ is confined to an annulus that has the origin
$J_1'=0$ within it. As $J_{1\,\rm res}'$ is reduced, the mean radius of the
annulus diminishes and
eventually the hole at its centre disappears. Then the dynamics plays out in
a disc that encompasses the origin and we need to move to coordinates that
are not singular there.

\begin{figure}
\centerline{\includegraphics[width=.7\hsize]{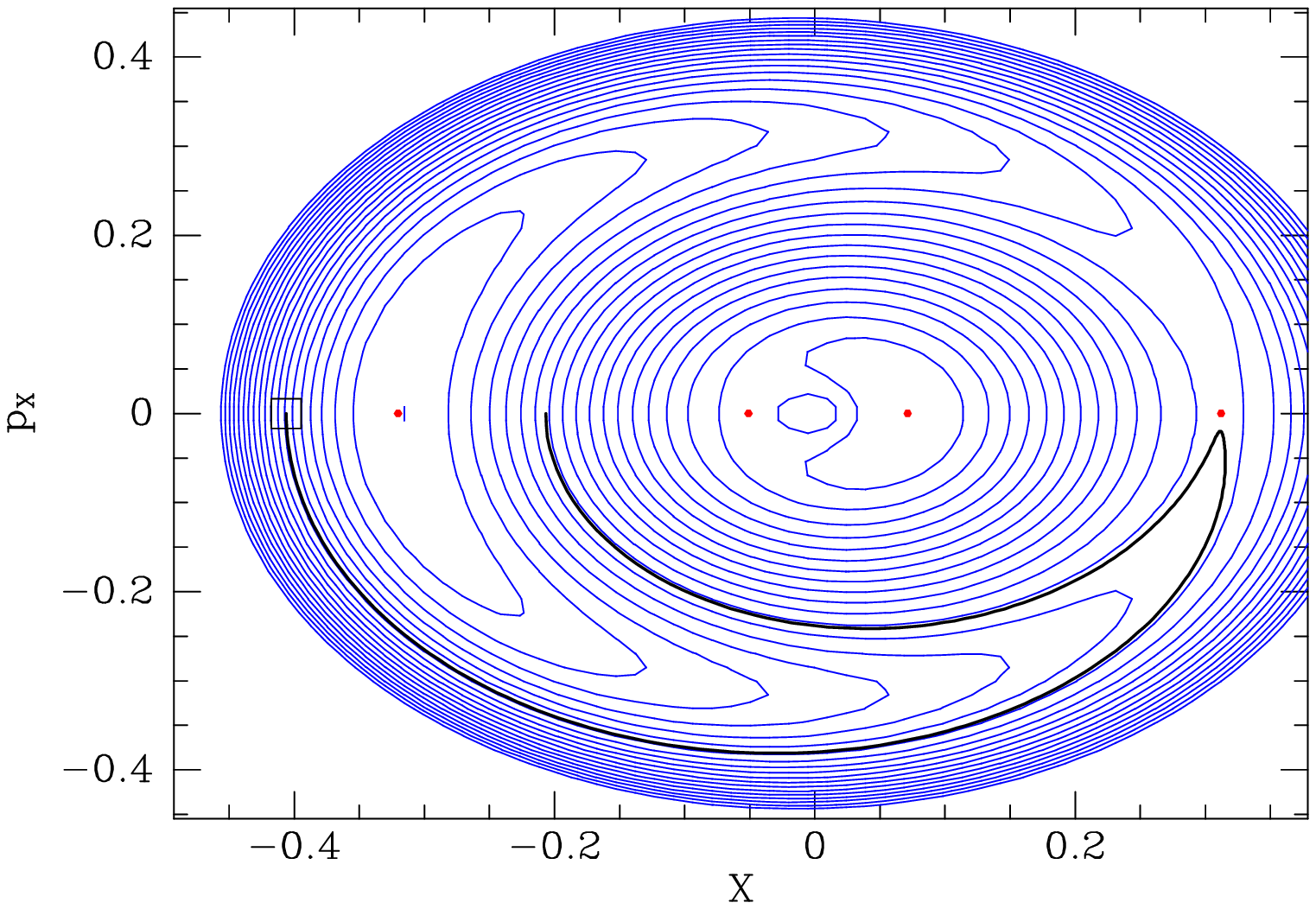}}%
\centerline{\includegraphics[width=.7\hsize]{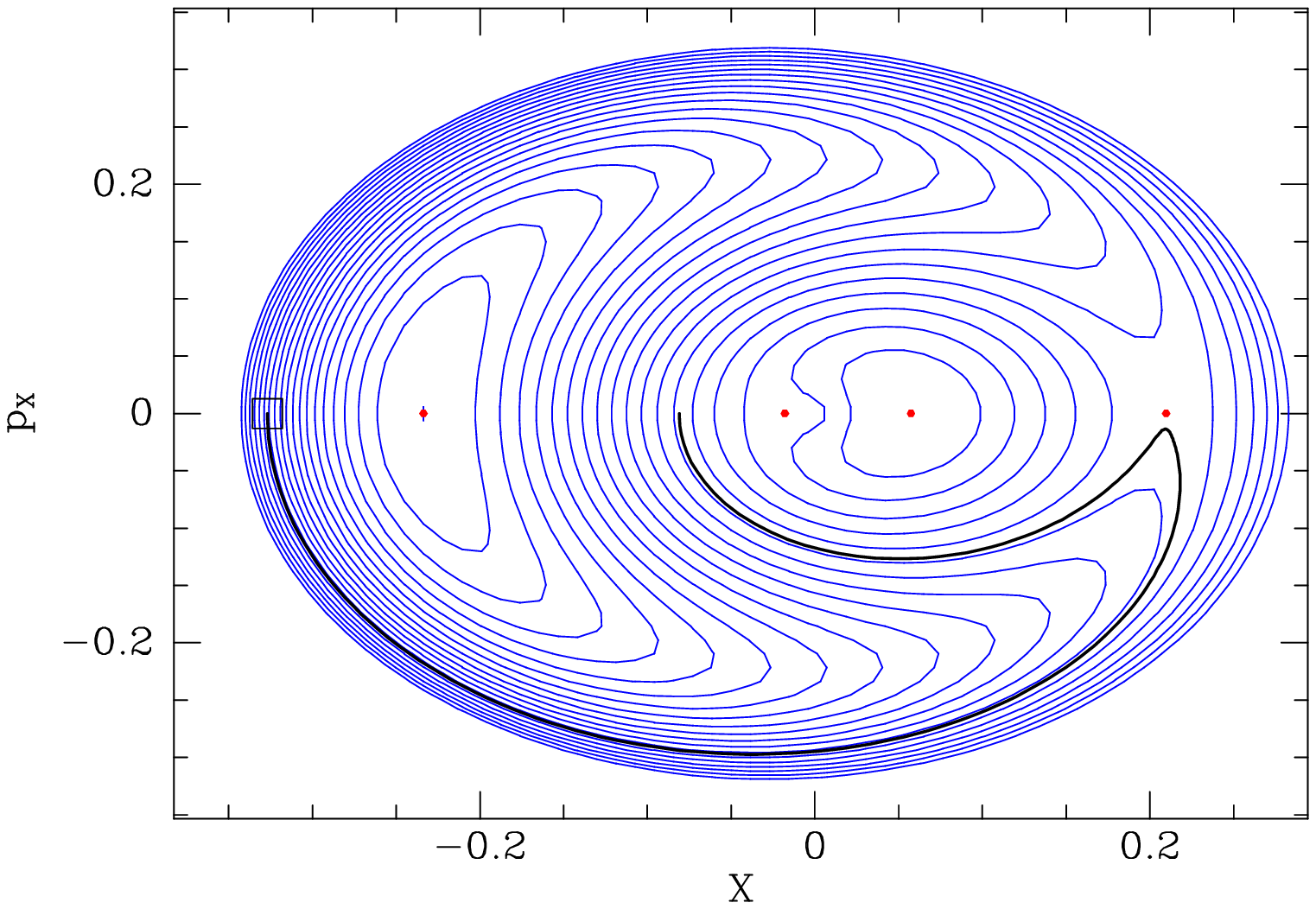}}%
\centerline{\includegraphics[width=.7\hsize]{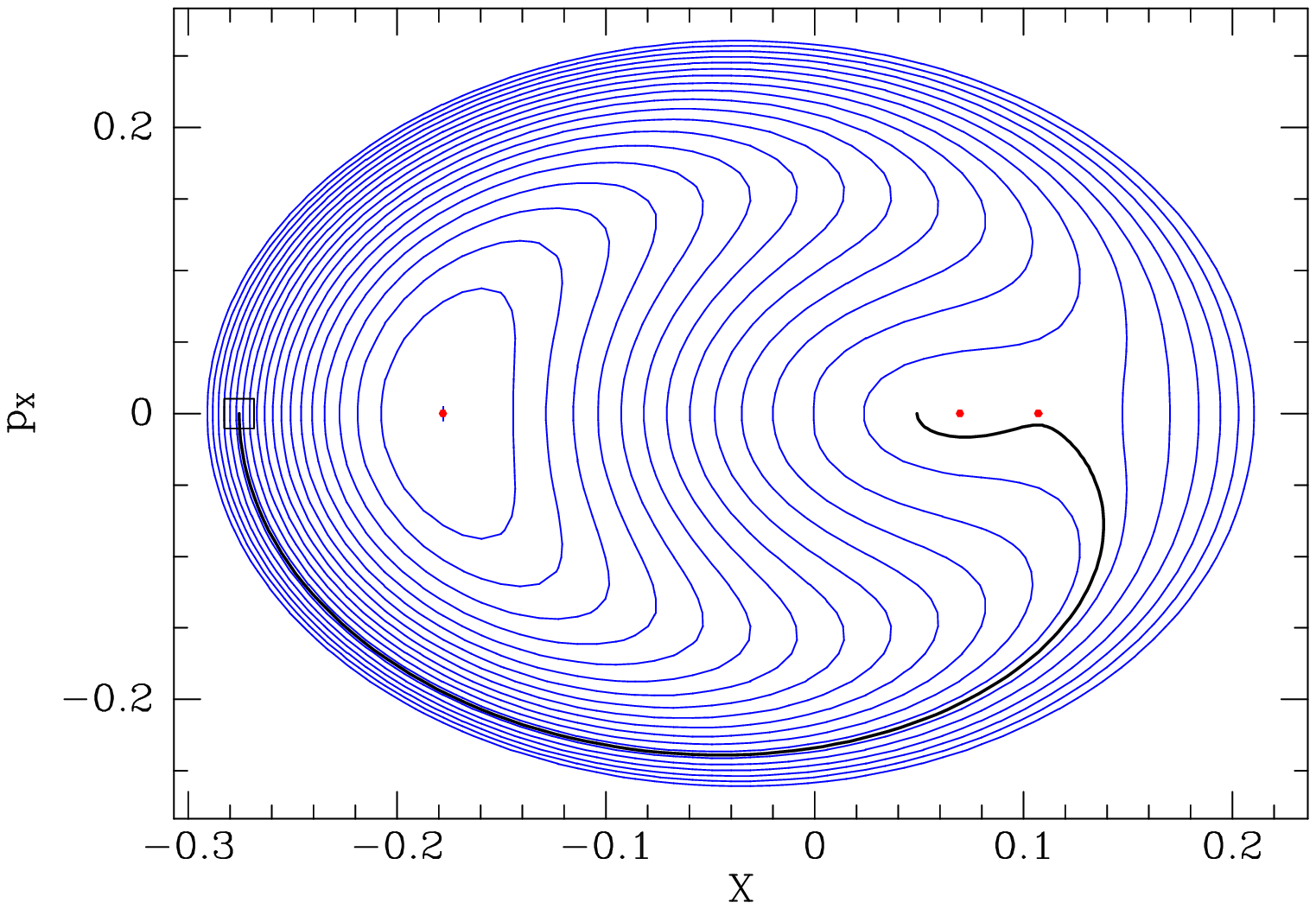}}%
\centerline{\includegraphics[width=.7\hsize]{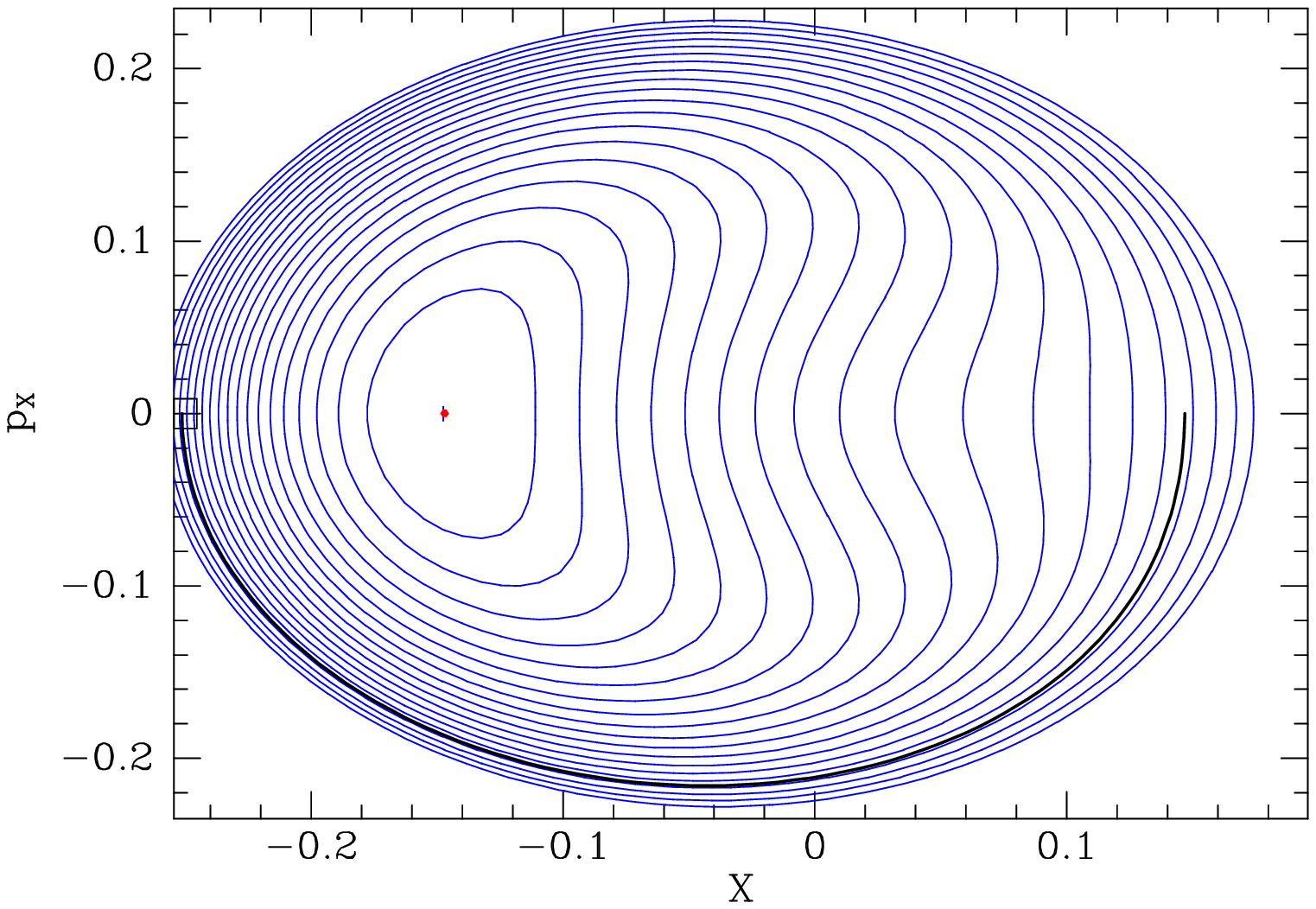}}%
\caption{Contours of constant $H(\theta_1',J_1')$ when $J_{1\,\rm res}'$
is diminished by factors of 2  from $J_{1\,\rm res}'=0.05$ (top) to
$0.00625$ (bottom) in the coordinates defined by
equations (\ref{eq:XpX}). Each panel corresponds to a rung of the ladder in
Fig.~\ref{fig:OLR_ladder}, with the upper panels corresponding to higher rungs
than the lower panels. $H(X,p_X)$ is  computed using quartics in
$\xi=\sqrt{2J_1'}$ to fit $H_0(J_1')$ and $h_N(J_1')$. The black contours are
trajectories obtained by integrating Hamilton's equation in the slow plane to
define the tori with the largest possible actions of libration.}\label{fig:plot_Heff}
\end{figure}

From the generating function
\[
S(X,\theta_1')=\fracj12X^2\tan\theta_1',
\]
we have
\begin{align}
p_X&={\p S\over\p X}=X\tan\theta_1'\cr
J_1'&={\p S\over\p\theta_1'}=\fracj12X^2\sec^2\theta_1',
\end{align}
so our new canonical coordinates are
\[\label{eq:XpX}
X=\xi\cos\theta_1';\quad p_X=\xi\sin\theta_1',
\]
where 
\[\label{eq:defsXi}
\xi\equiv\sqrt{2J_1'}.
\]
Fig.~\ref{fig:plot_Heff} shows contours of the Hamiltonian that controls the
dynamics in the $(J_1',\theta_1')$ plane at four values of $J_3'$. This is the
Hamiltonian that controls the motion of stars along the corresponding rung of
the ladder in Fig.~\ref{fig:OLR_ladder}. In Fig.~\ref{fig:plot_Heff} the values
of $J_r$ at which the rung cuts the ladder's centre line 
drop by steps of a factor 2 from $J_r=0.05$ in the top panel to
$J_r=0.00625$ in the bottom panel. 

\begin{figure}
\includegraphics[width=\hsize]{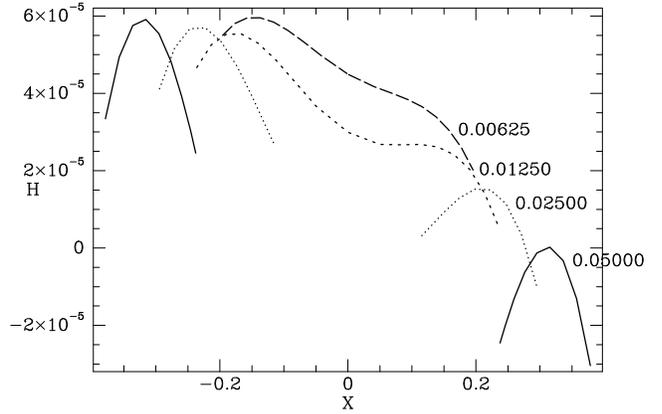}
 \caption{The values of $H$ across the centres of each panel of
Fig.~\ref{fig:plot_Heff}. Each panel corresponds to a rung of the ladder
shown in Fig.~\ref{fig:OLR_ladder} and the corresponding  curve is labelled by the value
of $J_r$ at which its rung crosses the ladder's centre line. 
For clarity the curve for $J_r=0.025$ is displaced upwards by
$10^{-5}$, that for $J_r=0.0125$ by $2\times10^{-5}$, etc.
}\label{fig:Xsect}
\end{figure}

In the top panel the annulus has mean radius $\sim0.3$ and is occupied by
crescent-shaped contours that enclose a stationary point marked by a red dot
at $X\simeq-0.3$. The outermost crescent almost touches itself at the stationary
point marked by a red dot on the right at $X\simeq0.3$.  Fig.~\ref{fig:Xsect}
shows the value of $H$ along the $X$ axes of the four panels of
Fig.~\ref{fig:plot_Heff}. We see that the stationary points on the left
Fig.~\ref{fig:plot_Heff} are mountain tops, while that stationary points on
the right of the upper panels are saddle points; in the top two panels of
Fig.~\ref{fig:plot_Heff} a lake bottom lies between
these features. As $J_{1\,\rm res}'$ falls, the lake becomes less deep, and
eventually its bottom annihilates with the saddle point, so one is left with
a mountain that has a steep left-hand face while allowing a fairly easy
ascent from the right.

This annihilation occurs at the value of $J_3'$ of the rung
that first touches the $J_\phi$ axis in Fig.~\ref{fig:OLR_ladder}. Rungs higher
up the ladder terminate at the edge of the ladder, leaving space between
their ends and the $J_\phi$ axis for orbits that circulate inside OLR.

The values of $H$ plotted in Figs.~\ref{fig:plot_Heff} and \ref{fig:Xsect}
were computed as follows. 

\begin{itemize}
\item The complete Hamiltonian was Fourier analysed on the perfectly resonant
torus $J_{1\,\rm res}'$, and on tori with slightly smaller and larger values
of $J_1'$.  As $J_r$ changed along this sequence, $J_\phi$ was adjusted to
hold constant $J_3'$. From the Fourier amplitude $H_0$ for vanishing
wavenumber and that, $h_N$, for the resonant wavenumber, the coefficients of
the terms appearing in the standard pendulum equation were determined.

\item The pendulum equation then yielded estimates of the smallest and
largest values of $J_1'$ reached by the pendulum when it librates with
maximum amplitude. The dashed red curves in Fig.~\ref{fig:OLR_ladder} show these
limiting values; they are evidently excessive so are reduced by a factor
$0.7$. If the smallest value, $J_{1\,\rm min}'$, was predicted to be
negative, it was replaced by a tiny positive value.

\item The Hamiltonian was then Fourier analysed on nine tori that had values
of $J_1'$ that spanned the range $(J_{1\,\rm min}',J_{1\,\rm max}')$ with
$J_\phi$ again adjusted to hold constant $J_3'$. The data for the
amplitudes $H_0(J_1')$ and $h_N(J_1')$ obtained in this way, which are
plotted in Fig.~\ref{fig:Xsect}, were fitted by
quartics in $\xi$ (eqn.~\ref{eq:defsXi}). In B18, by contrast, these amplitudes
were fitted by quadratics in 
\[
\Delta\equiv J_1'-J_{1\,\rm res}'.
\]
 The use of
$\xi$ rather than $\Delta$ as the independent variable is partly motivated by
equation (\ref{eq:XpX}), and partly by the consideration that as $J_1'\to0$,
smoothness of $H(X,p_X)$ near the origin requires that $h_N\sim\xi$ because
it is a dipole amplitude. The need for quartics is indicated by the data  in
Fig.~\ref{fig:Xsect} for $J_r=0.0125$ shown by the short-dashed curve: in
addition to a local maximum at $X\sim-1.9$ the curve has a point of
inflection near the origin. This structure cannot be fitted by a quadratic.

\item Given that $h_N(0)=0$, the quartic fitted to the data for $h_N$
should have no constant term. The quartic fitted to the data for $H_0$ can be
similarly specialised since we know that $0=\p H_0/\p J_1'$ at $J_{1\,\rm
res}'$.

\end{itemize}

The quartic in $\xi$ is fitted to data that  covers a region only slightly
bigger than the  annuli in Fig.~\ref{fig:plot_Heff} within which orbits can be
trapped. Consequently, no significance should be attached to the quartic's values well
outside this region. In particular, the bumps in the bottom of the lake that
feature in the top two panels of Fig.~\ref{fig:plot_Heff} are of no
significance.

For the effective Hamiltonians plotted in the upper three panels of
Fig.~\ref{fig:plot_Heff} there are two critical values of $H$ to determine.
The first is the value associated with the red dot at negative $X$ around
which the contours of trapped orbits circulate; this is the value of $I$
associated with vanishing action of libration $\cJ$ (B18 eqn.~37). In the
case of the OLR, this value is a maximum (Fig~\ref{fig:Xsect}). The other
critical value is the value of $H$ at the saddle point along the positive $X$
axis, which divides trapped from untrapped orbits.  These values are computed
by finding the roots of the cubic equations obtained by differentiating
\[
H(\theta_1',\xi)=H_0(\xi)+2h_N(\xi)\cos(\theta_1')
\]
 with respect to $\xi$ at $\theta_1'=0$ and $\pi$. Only non-negative real
roots need be considered, but there can be as many as six such roots. Roots
that lie far from $\xi_{\rm res}$ are discarded, and when three roots survive
this cut, one chooses a saddle point and a maximum if $G<0$ or a minimum
when $G>0$.

The effective Hamiltonian that is plotted in the bottom panel of
Fig.~\ref{fig:plot_Heff} has only one critical value, namely the value of its
maximum, which characterises the orbit that has vanishing action of
libration. A value is adopted as a minimum that creates orbits that
effectively circulate around the origin of the $(\theta_1',J_1')$ plane. This
arbitrarily chosen value sets the upper boundary of the lower rungs in
Fig.~\ref{fig:OLR_ladder}. 

The algorithm of B18 for reconstructing trapped orbits is as follows.
At any given value of $\theta_1'$, one finds the roots $\Delta_\pm$ of the
quadratic in $\Delta$
\[
I=H_0(\Delta)+2h_N(\Delta)\cos\theta_1'.
\]
In terms of the top panel of Fig.~\ref{fig:plot_Heff}, this amounts to finding
the points at which a ray from the origin cuts the crescent-shaped contour
$H=I$. Once these points are known for all rays that cut the contour, it is
straightforward to evaluate the integral
\[
\cJ=\pi^{-1}\int\d\theta_1'(\Delta_+-\Delta_-)
\]
 and a similar integral
that yields the conjugate variable $\vartheta(\theta_1')$.

It is easy to see that this algorithm may fail if the structure of the
Hamiltonian is like those shown in the bottom two panels of
Fig.~\ref{fig:plot_Heff} because for large values of $\cJ$ each ray cuts the
contour $H=I$ only once. Instead we find where the $X$ axis cuts the contour
$H=I$ on the extreme left, and from there integrate Hamilton's equations
\[
{\d X\over\d\tau}={\p H\over\p p_X},\quad{\d p_X\over\d\tau}=-{\p H\over\p X}
\]
until $p_X$ changes sign, at $\tau_1$, say. The black half-contours in
Fig.~\ref{fig:plot_Heff} show results of such integrations. Values of $X,p_X$
and $\tau$ are stored at each step. When the integration finishes, the frequency of
libration is  $\Omega_\ell=\pi/\tau_1$ and the stored times can be converted
to true angles through  $\vartheta=\Omega_\ell\tau$.
The derivatives required for Hamilton's equations are available analytically:
\begin{align}
{\p H\over\p p_X}
%={\p H\over\p\theta_1'}{\p\theta_1'\over\p p_X}+{\p H\over\p\xi}{\p\xi\over\p p_X}\cr
&=2h_N\sin\theta_1'{X\over\xi^2}+{\p H\over\p\xi}{p_X\over\xi}\cr
{\p H\over\p X}&=-2h_N\sin\theta_1'{p_X\over\xi^2}+{\p H\over\p\xi}{X\over\xi}.
\end{align}

\begin{figure}
\includegraphics[width=\hsize]{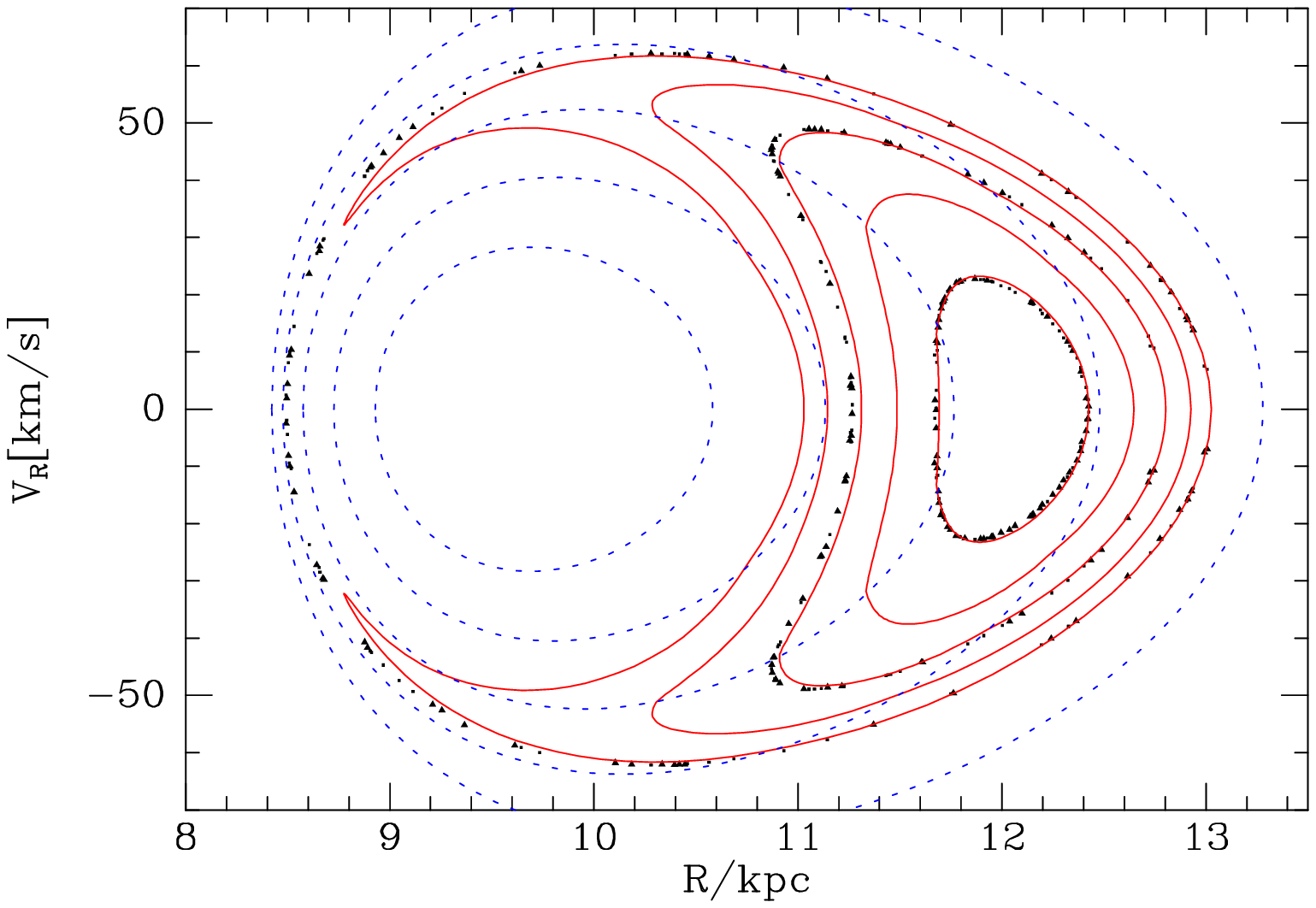}
\includegraphics[width=\hsize]{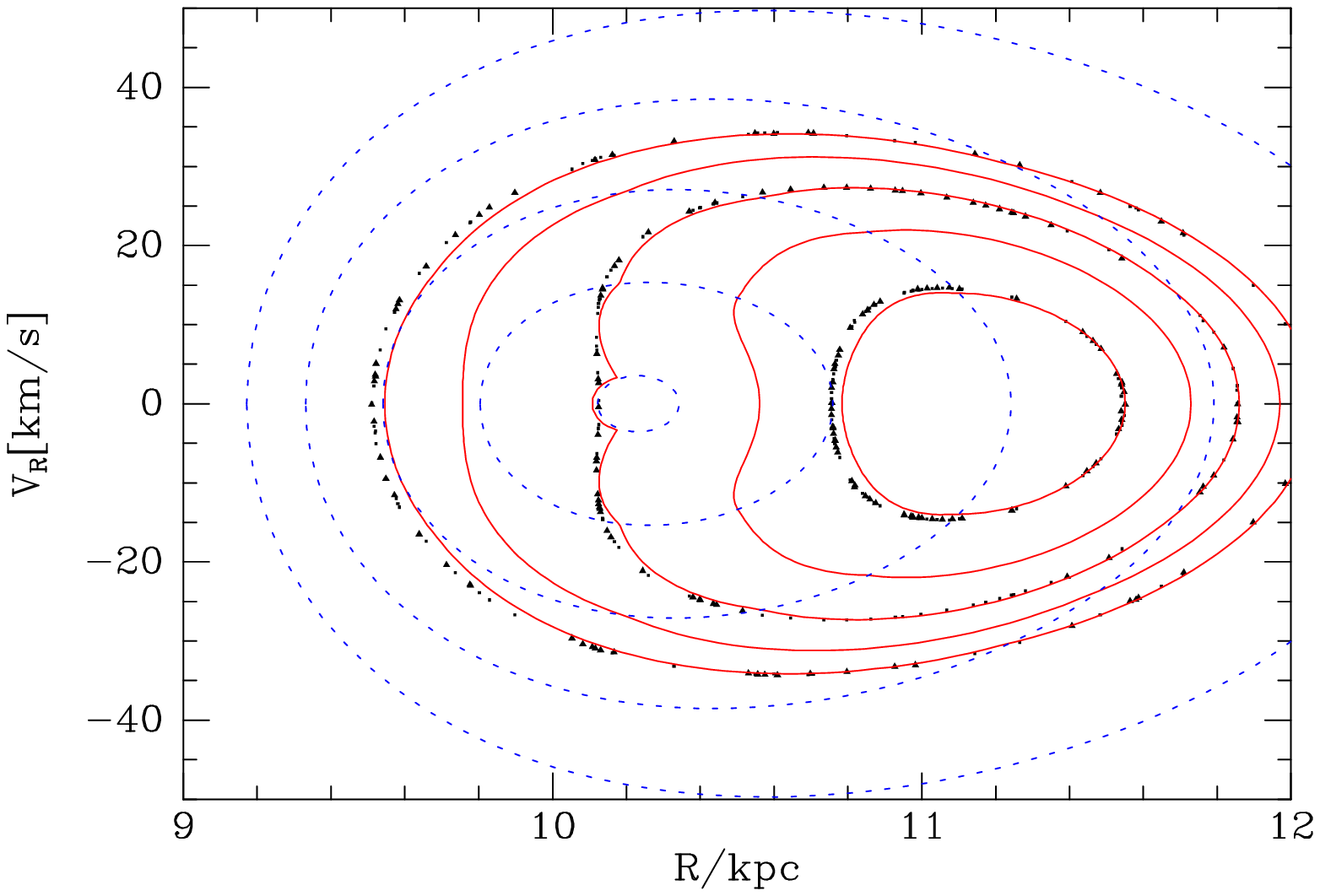}
\caption{Orbits trapped at OLR in surfaces of section for $J_r=0.05$ (upper) and
$J_r=0.00625$ (lower). The full red curves are computed by resonant
perturbation theory while the black dots are obtained by Runge-Kutta
integration of the full equations of motion. The dashed blue curves are cross
sections of some of the axisymmetric tori that underlie the perturbation
theory. }\label{fig:OLRSoS}
\end{figure}

\section{Orbits trapped at OLR}\label{sec:OLR}

Each panel of Fig.~\ref{fig:OLRSoS} shows cross sections through five orbits
trapped at OLR. The full red curves are computed from an object returned by
{\tt resTorus\_L} and calling its method {\tt SOS} for five values of $I$
between $I_{\rm min}$ and $I_{\rm max}$. The black dots are consequents obtained by
integrating the full equations of motion with a fourth-order Runge-Kutta
routine starting from one point on three of the full curves. The dashed blue
curves are cross sections through some of the axisymmetric tori that underlie the
computation and are drawn by the method {\tt SOS} of the corresponding {\tt
Torus} object. The fit between the dots and the full curves is almost perfect except
where the curves cross the $R$ axis on the left. The worst errors occur
between the smallest and next smallest blue curves in the lower panel, and
reflect unresolved issues regarding interpolating between
tori that require a point transformation \citep[Appendix A3 of][]{JJBPJM16}.

The upper panel of Fig.~\ref{fig:OLRSoS} corresponds to the top panel of
Fig.~\ref{fig:plot_Heff}. At this value of $J_3'=J_\phi-2J_r$ the range in
$J_r$ over which orbits librate has a lower as well as an upper bound. Orbits
with small $J_r$ circulate in $\theta_1'$ too fast to be trapped, while
orbits with largest $J_r$ circulate too fast the opposite direction to be
trapped. In Fig.~\ref{fig:OLRSoS} the horns of the crescents reach round to
touch each enclosing an island of blue dashed ovals associated with
circulating orbits with small $J_r$. In the top panel of
Fig.~\ref{fig:plot_Heff} these circulating orbits move along ovals that cut
the positive $X$ axis to the left of the saddle point; they tour the lake.

The lower panel of Fig.~\ref{fig:OLRSoS} corresponds to the bottom panel of
Fig.~\ref{fig:plot_Heff}. At this small value of $J_3'$, $\theta_1'$
circulates so slowly in the axisymmetric potential that the bar traps the
orbit, even at negligible
radial action. Hence only orbits with large $J_r$ circulate, so in the bottom panel
of Fig.~\ref{fig:plot_Heff} there are no small ovals centred on the origin,
and in Fig.~\ref{fig:OLRSoS} the origin lies within the domain of red invariant
curves that encircle $R\sim11.3\kpc$. 

In the bottom two panels of Fig.~\ref{fig:plot_Heff}  the distinction
between trapped and untrapped orbits has become unclear. The smallest ovals
in Fig.~\ref{fig:plot_Heff} describe orbits that keep close to an unperturbed
orbit with a particular value of $J_r$ (marked by the red dot) that fails to
precess as it would in the underlying axisymmetric potential. As one moves
out to larger ovals, the corresponding orbit cycles through a range of values
of $J_r$ that at first widens until it embraces the circular orbit $J_r=0$.
Later the range gradually contracts while its centre shifts to high values
of $J_r$; this contraction is signalled in Fig.~\ref{fig:plot_Heff} by the
outer ovals becoming more nearly centred on the origin.
During the widening stage, the orbit's major axis oscillates with increasing
amplitude synchronously with variation in its eccentricity. During the
contracting stage, the major axis precesses with growing steadiness as the
orbit becomes more eccentric. Thus the transition from absolute confinement
(at the red dot) to motion very close to that expected in the absence of the
bar, is continuous.

As a consequence of this continuity, the lower rungs of the ladder of
Fig.~\ref{fig:OLR_ladder} do not have unambiguous upper ends. The upper ends of
the higher rungs are, by contrast, clearly defined by the value of $\cJ$ at
which circulation sets in. In Fig.~\ref{fig:OLR_ladder} this ambiguity manifests
itself in a slight kink in the ladder's top boundary, which has been set
arbitrarily.

\begin{figure}
\centerline{\includegraphics[width=.7\hsize]{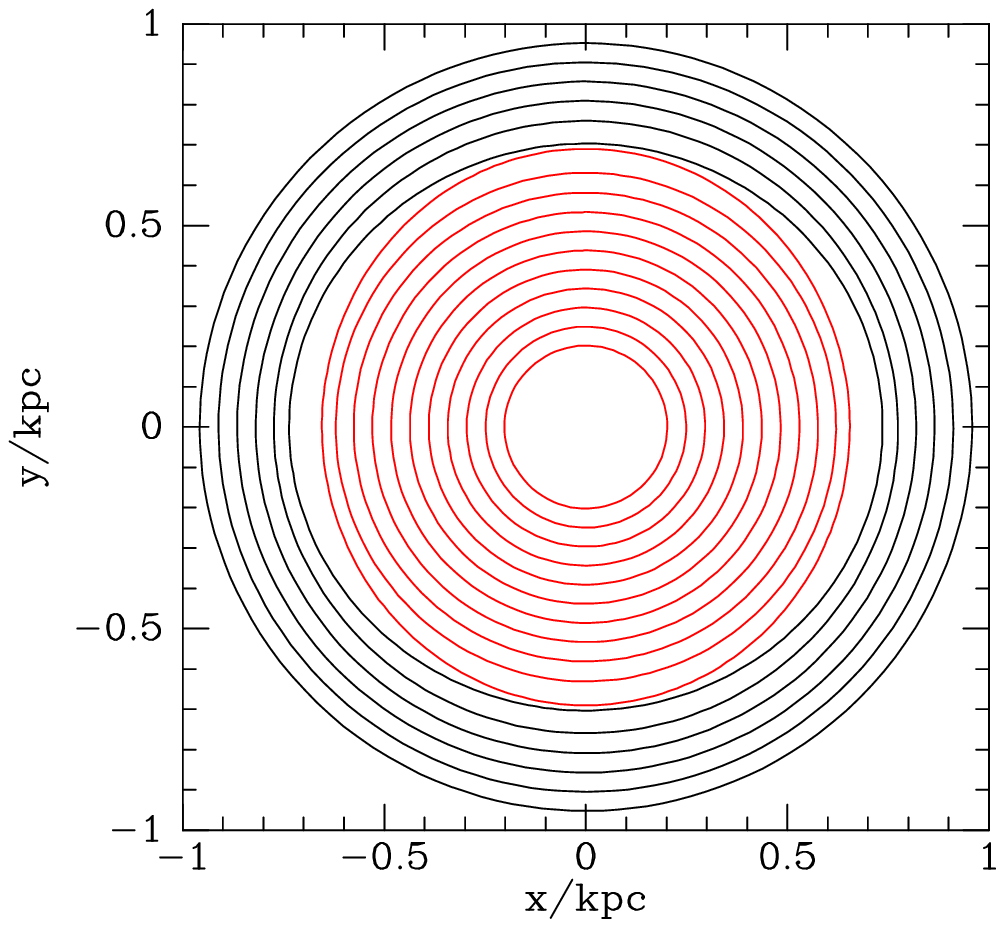}}
\centerline{\includegraphics[width=.8\hsize]{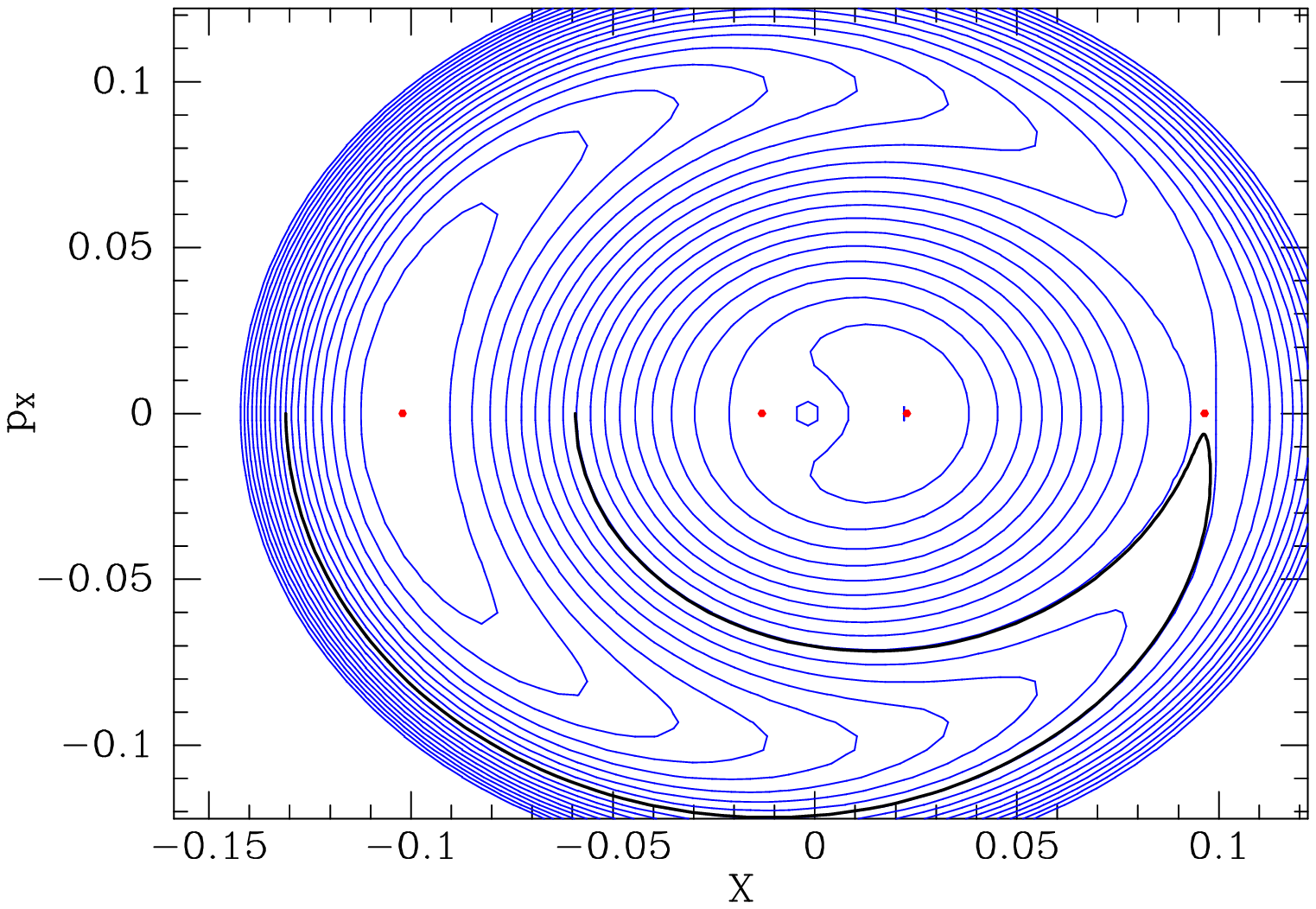}}
\centerline{\includegraphics[width=.8\hsize]{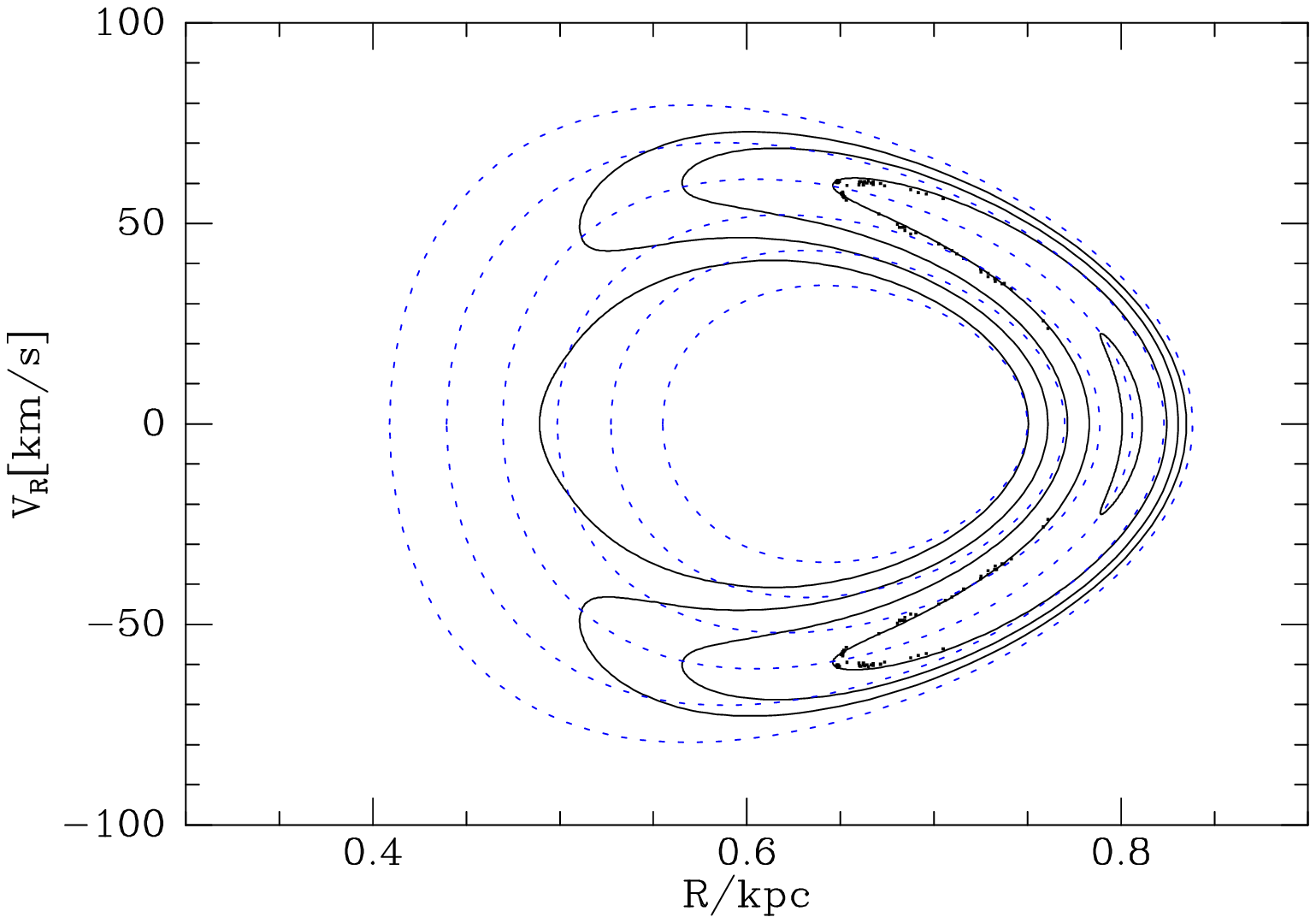}}
\caption{Trapping at the ILR of a weak bar: {\tt Bstrength=0.005} The lower
two panels are for $J_r=0.005$.
}\label{fig:weak}
\end{figure}

\section{Trapping at the ILR}\label{sec:ILR}

The non-axisymmetric part of the potential becomes more prominent relative to
the axisymmetric part as one moves towards the Galactic centre, with the
consequence that in a given bar perturbation theory works less well for
orbits trapped at the ILR than at the OLR. This being so,
results are presented for bars that are weaker than that studied at the OLR. In fact we
will use B18 bar with amplitude multiplied  by a factor
${\tt Bstrength}<1$.

The ILR is defined by the vector $\vN=(1,0,-2)$.  Whereas at the OLR, an
axisymmetric, perfectly resonant torus yields $G<0$ and $\psi_\vN=\pi$, at
the ILR $G>0$ and $\psi_\vN=0$. Libration is still around $\theta_1'=\pi$,
but now stars move around a minimum rather than a maximum in $H$. 

Fig.~\ref{fig:weak} shows the orbital structure of a very weak bar: ${\tt
Bstrength=0.005}$.  The top panel shows closed orbits. Two orbit families are
evident.  At radii larger than $\sim0.7\kpc$ the orbits belong to the $x_1$
family in the notation of \cite{ContopoulosP1980}, while interior to
$0.7\kpc$ the obits (shown in red) belong to the $x_2$ family. The centre panel of
Fig.~\ref{fig:weak} shows contours of the effective Hamiltonian in the slow
plane analogous to Fig.~\ref{fig:plot_Heff}. The structure is qualitatively
the same as that in the top panel of Fig.~\ref{fig:plot_Heff}. On the left,
crescent shaped contours are followed by orbits that are trapped, while on
the far right a saddle point sets the boundary between trapped and untrapped
orbits. The only difference between this situation and that at the OLR is that
now we have a lake rather than a mountain on the left and a hill rather than
a lake near the centre. 

\begin{figure}
\centerline{\includegraphics[width=.7\hsize]{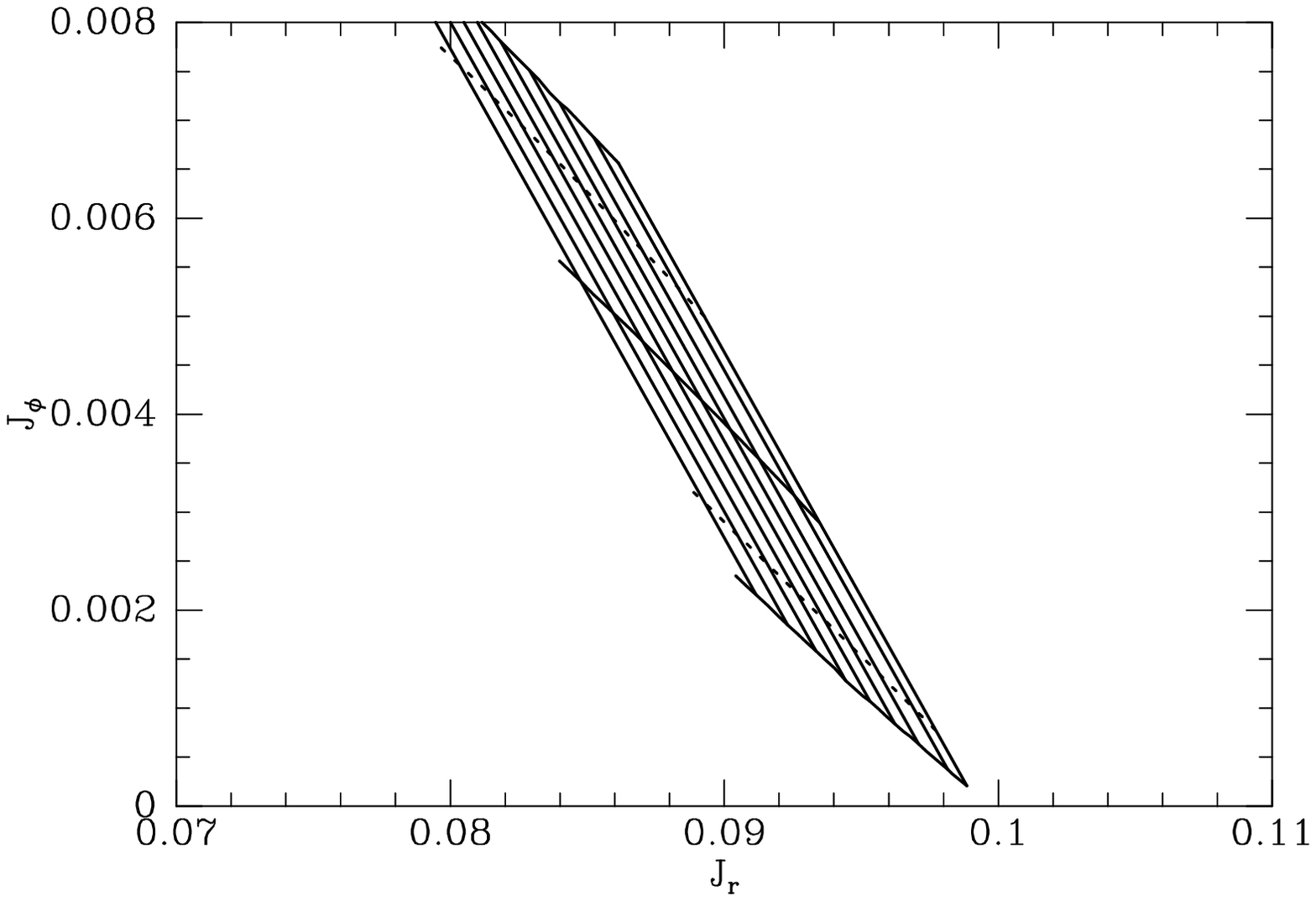}}
\centerline{\includegraphics[width=.7\hsize]{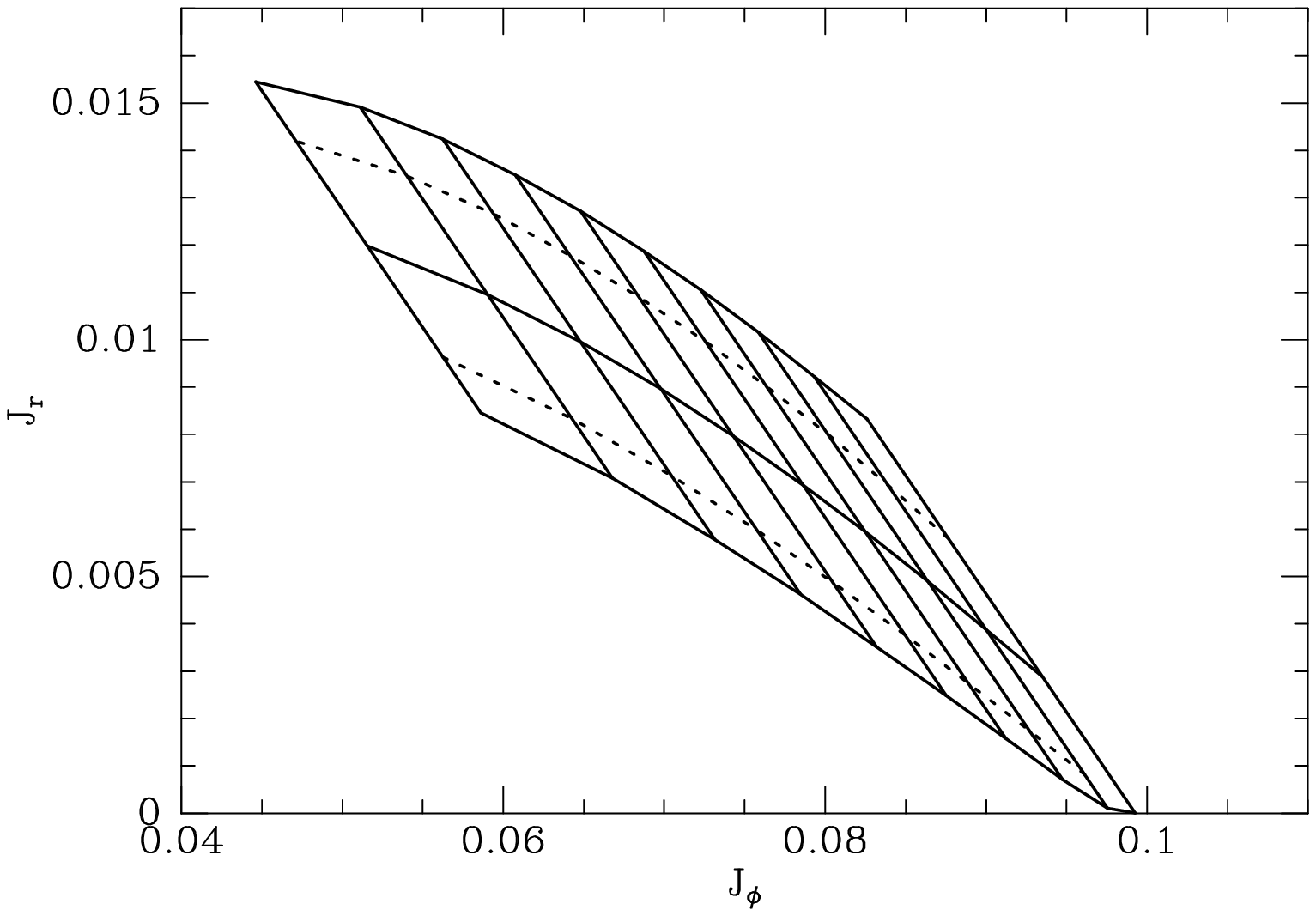}}
\centerline{\includegraphics[width=.7\hsize]{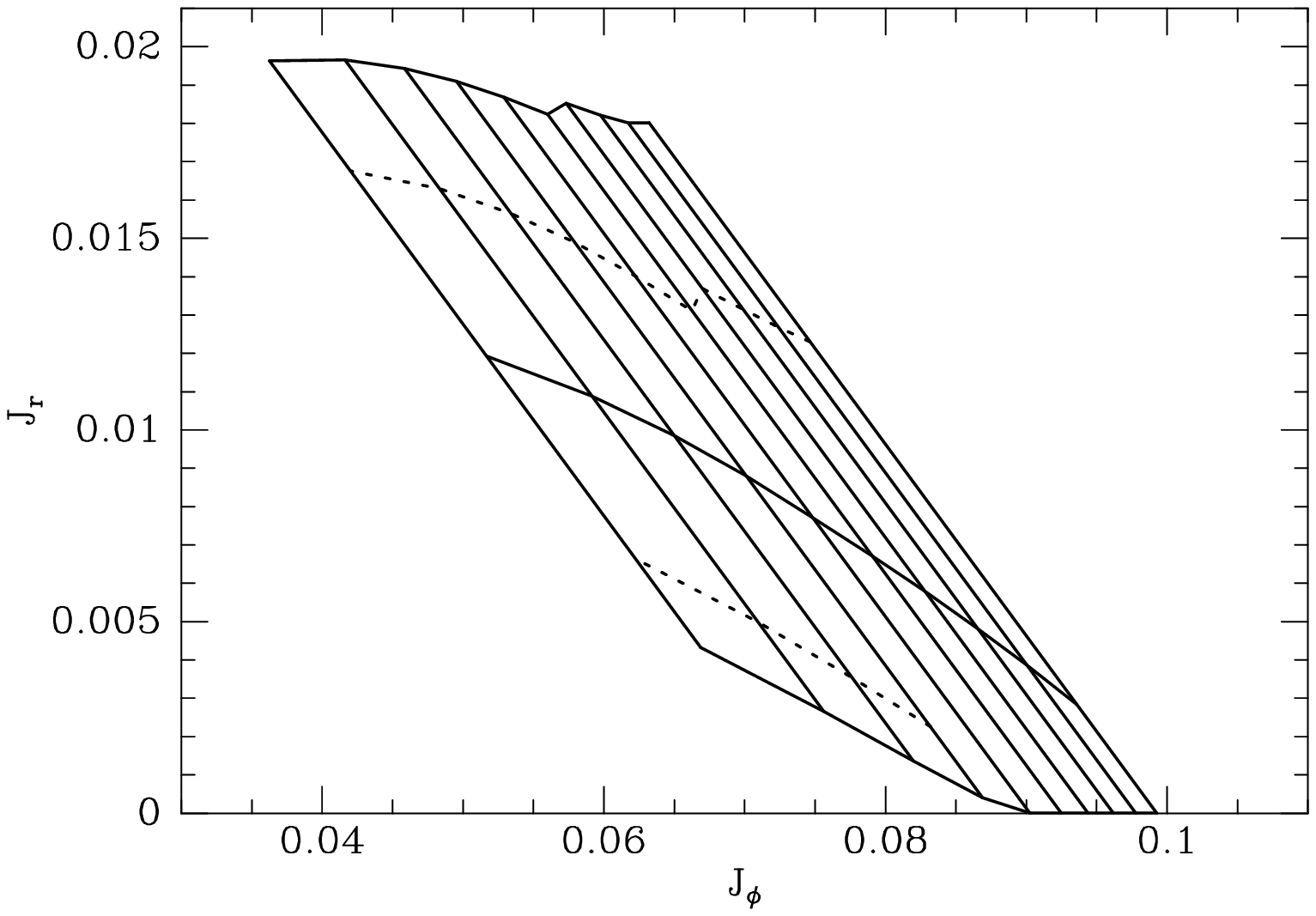}}
\centerline{\includegraphics[width=.7\hsize]{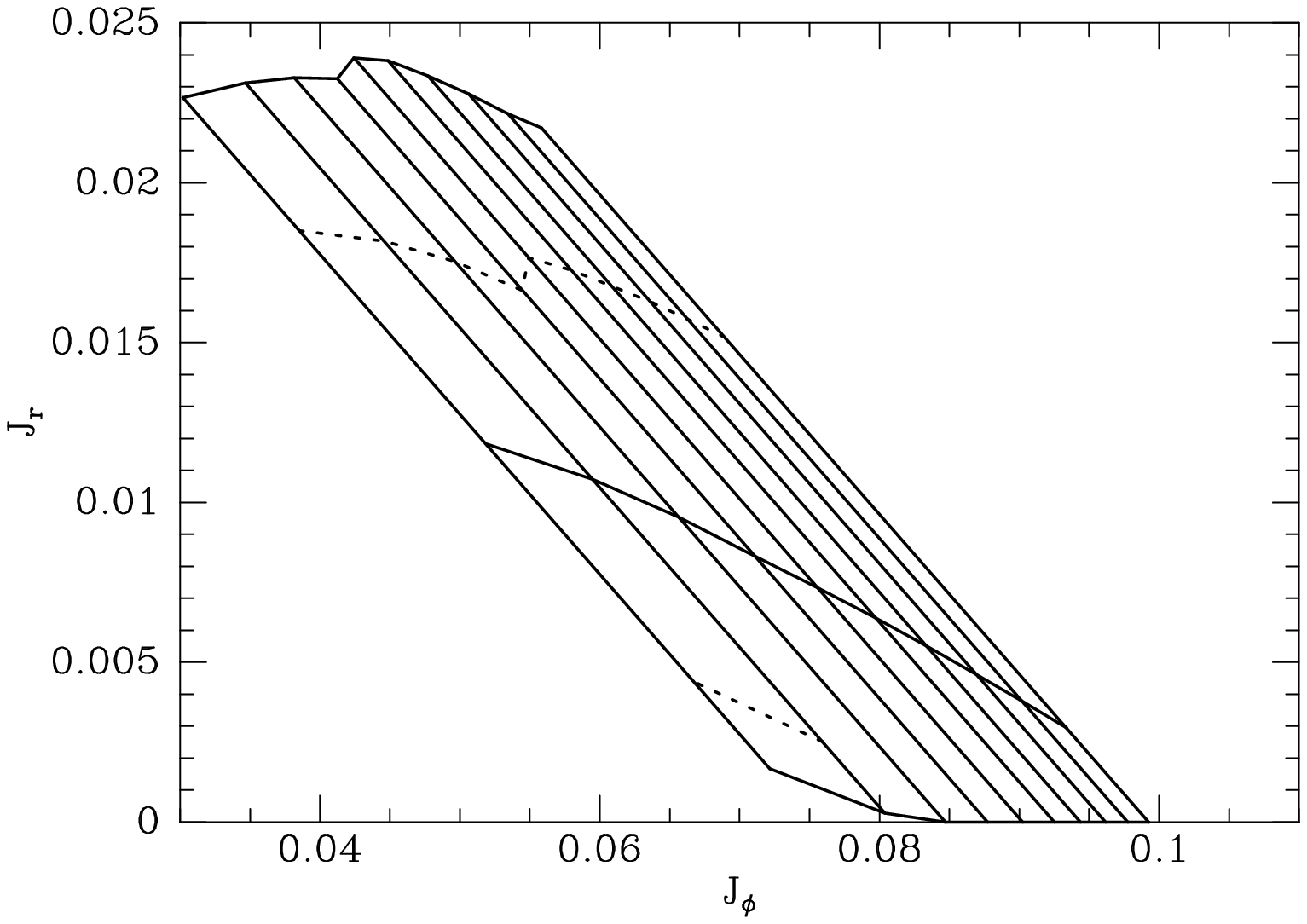}}
\caption{Orbits trapped at the ILR in bars of increasing strength: from top
to bottom $Bs=0.005$, $0.015$, $0.05$ and $0.1$.}\label{fig:ILR_ladder}
\end{figure}

Fig.~\ref{fig:ILR_ladder} shows the $(J_\phi,J_r)$ plane in the vicinity of
the ILR for bars with strengths that increase from top to bottom from $0.005$
to $0.1$.  The ladder in the bottom panel includes tori
with $J_r/J_\phi$ as large as $0.77$ even though the bar has only a tenth of
the strength of the B18 bar, and less than a tenth the strength of the S15
bar. Tori with larger ratios $J_r/J_\phi$ are costly to compute. Moreover, we
shall find that even in this bar trapped orbits are
significantly stochastic and are thus pushing the concept of an invariant
torus to its limit. For this reason in this section we restrict discussion to
${\tt Bstrength}\le0.1$.

Like the OLR's ladder in Fig.~\ref{fig:OLR_ladder}, the ladders in
Fig.~\ref{fig:ILR_ladder}
slope from lower right to upper left, but their rungs slope in the same sense
as the ladder
rather than in the opposite sense.
Again the higher rungs finish on the lower boundary of the ladder, leaving
room between that point and the $J_\phi$ axis for orbits to circulate inside
the ILR, while the lower rungs reach the $J_\phi$ axis.  As the strength of the
bar grows, the ladder becomes wider and area of the $(J_\phi,J_r)$ plane
available to orbits that circulate inside the ILR shrinks dramatically. As
the ladder widens, the eccentricity of orbits along its upper edge increases
and computing their tori becomes increasingly costly. 

The ladders of Fig.~\ref{fig:ILR_ladder} have marked kinks in their upper
boundaries at the last rung to start from the $J_\phi$ axis. These kinks
reflect the fact that that once the saddle point has disappeared from the
Hamiltonian in the slow plane, the transition from libration to circulation
occurs continuously rather than at a well defined value of $\cJ$, so the
upper ends of rungs have to be set arbitrarily.

\begin{figure}
\centerline{\includegraphics[width=.7\hsize]{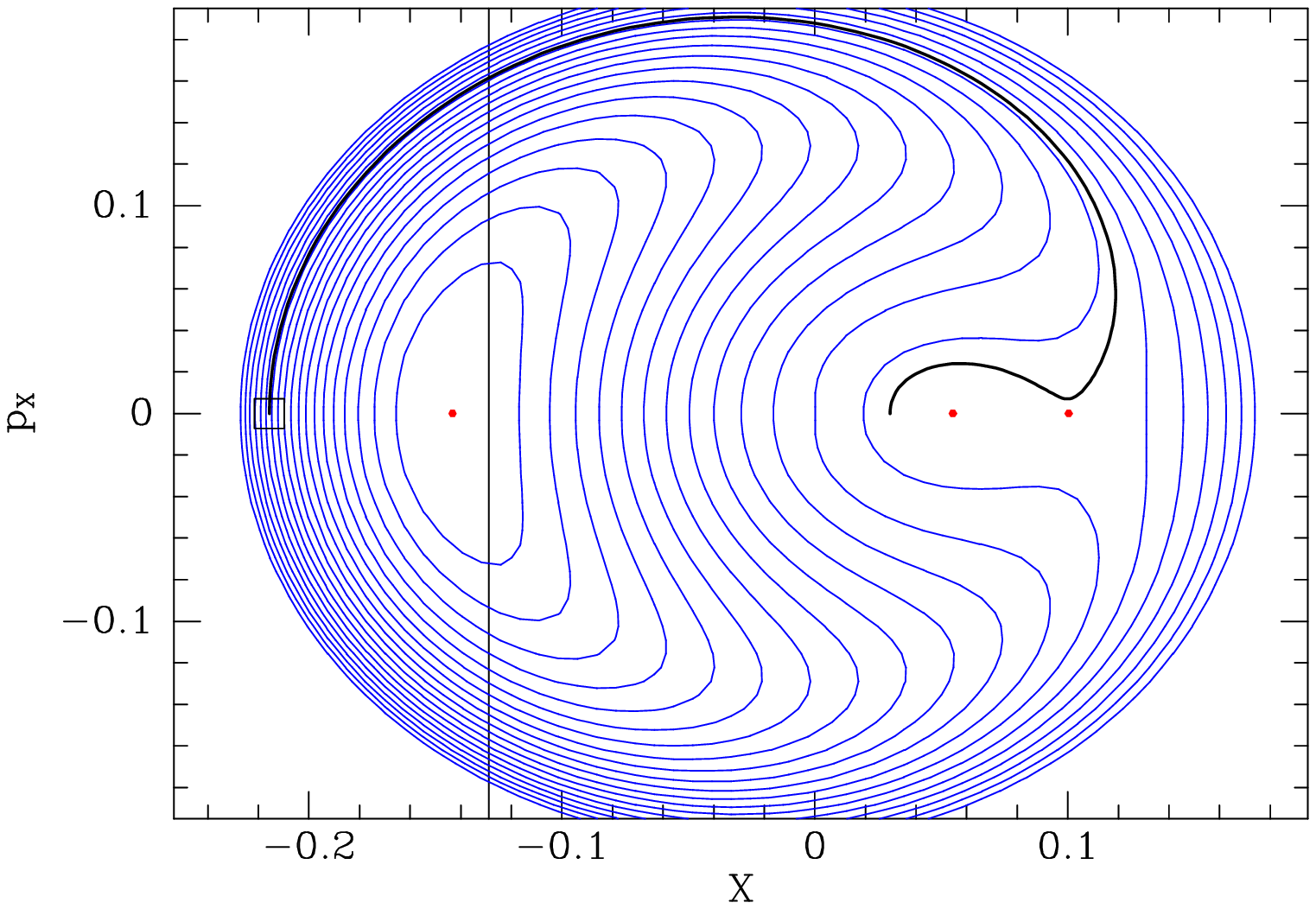}}
\centerline{\includegraphics[width=.7\hsize]{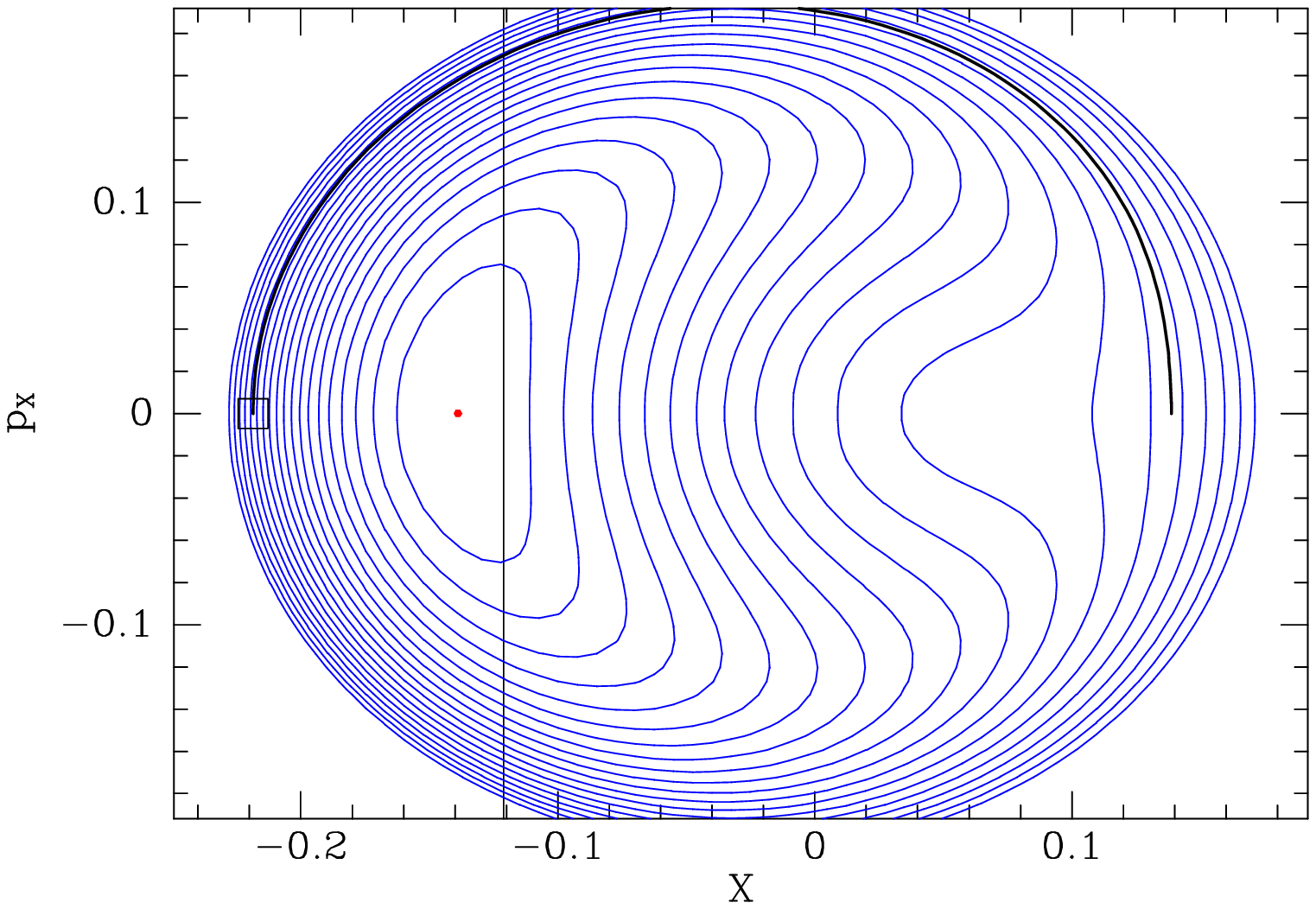}}
\centerline{\includegraphics[width=.7\hsize]{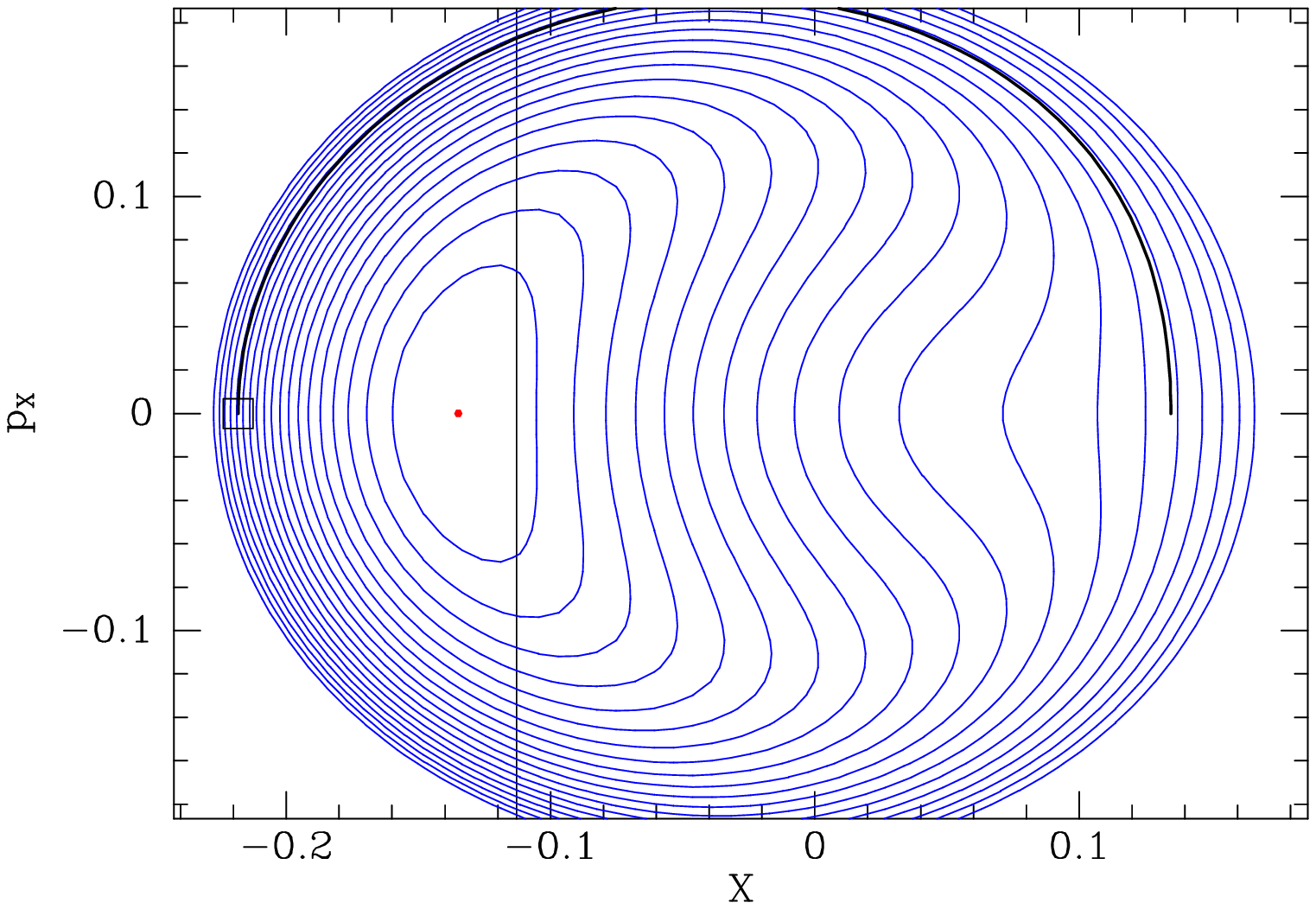}}
\caption{The Hamiltonian that governs the dynamics  of orbits near the  ILR
of a bar with strength $Sb=0.1$.  From top to bottom these Hamiltonians are
for the rungs of the ladder in the bottom panel of Fig.~\ref{fig:ILR_ladder}
that cut the ladder's centre line at $J_r=0.007$
$0.008$ and $0.009$.}\label{fig:ILRHeff}
\end{figure}

\begin{figure}
\centerline{\includegraphics[width=.9\hsize]{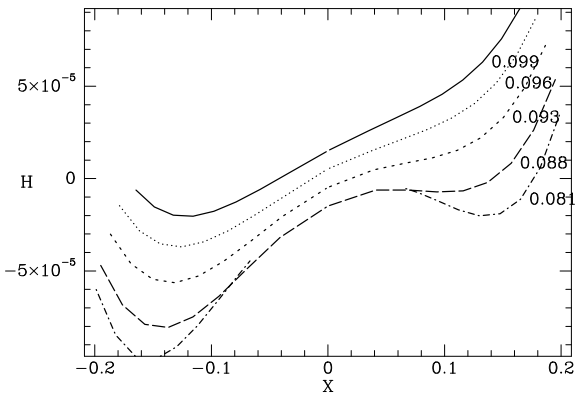}}
\caption{Effective potentials along the $X$ axis for trapping orbits at the
ILR of a bar of strength $Bs=0.1$.}\label{fig:ILR_HeffX}
\end{figure}

Fig.~\ref{fig:ILRHeff} is the analogue of Fig.~\ref{fig:plot_Heff} for the
ILR of a bar of strength $0.1$. 
The similarity between Fig.~\ref{fig:ILRHeff}
and the bottom two panels of Fig.~\ref{fig:plot_Heff} is striking. In the
lowest panels of both figures there is a continuous transition between
trapped and untrapped orbits. At both the ILR and the OLR, a clear
distinction emerges where the ladder's lower edge reaches the $J_\phi$ axis.
In Fig.~\ref{fig:ILRHeff} the rung from this junction corresponds to the
middle panel, where the hill is annihilating with the saddle point.
Fig.~\ref{fig:ILR_HeffX} makes this evolution clearer by showing
 % for ${\tt Bstrength}=0.1$
 the value of $H$ along the $X$ axis for several values of
$J_3'=J_\phi+2J_r$. The curves for larger values of $J_3'$ (low on the
ladder) have a global minimum on the left and slope steadily upwards on the
right. Around $J_3'=0.088$ the curve develops a point of inflection which
matures into a hill and a depression (saddle point) as $J_3'$ decreases
further (i.e., one moves up the ladder). 

\begin{figure*}
\centerline{\includegraphics[width=.33\hsize]{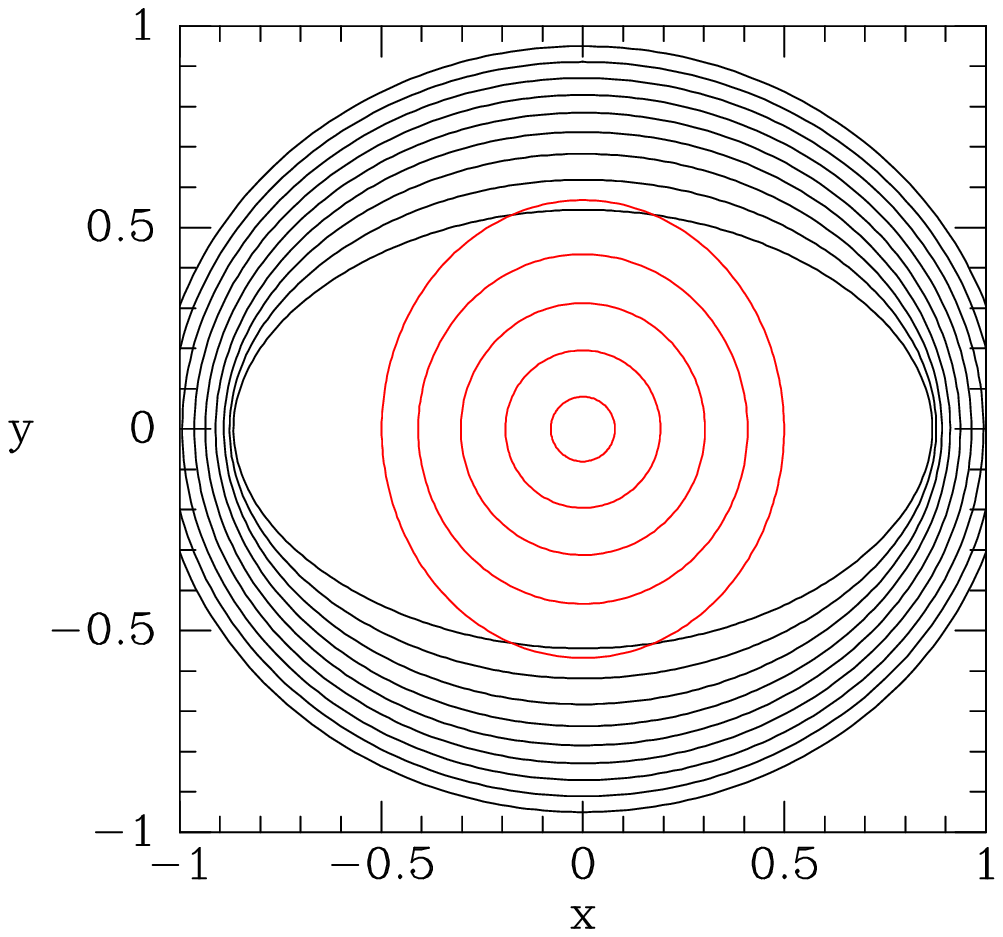}
\includegraphics[width=.33\hsize]{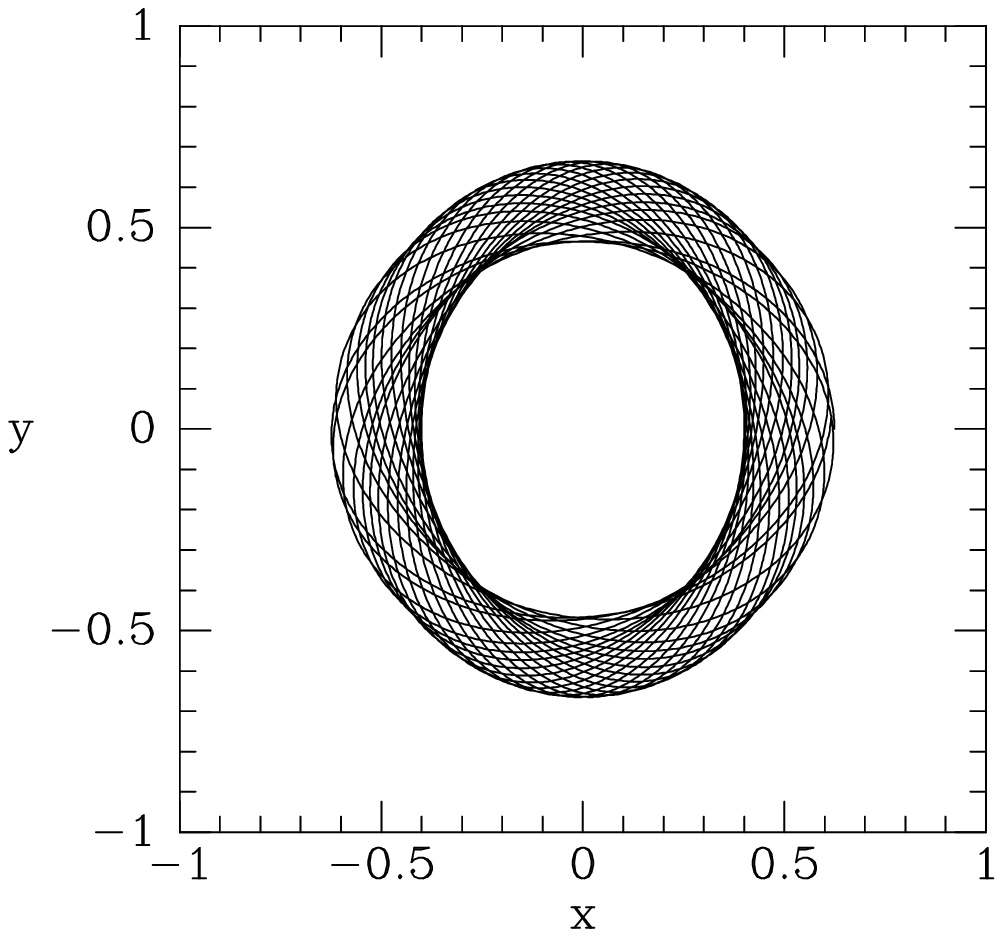}
\includegraphics[width=.33\hsize]{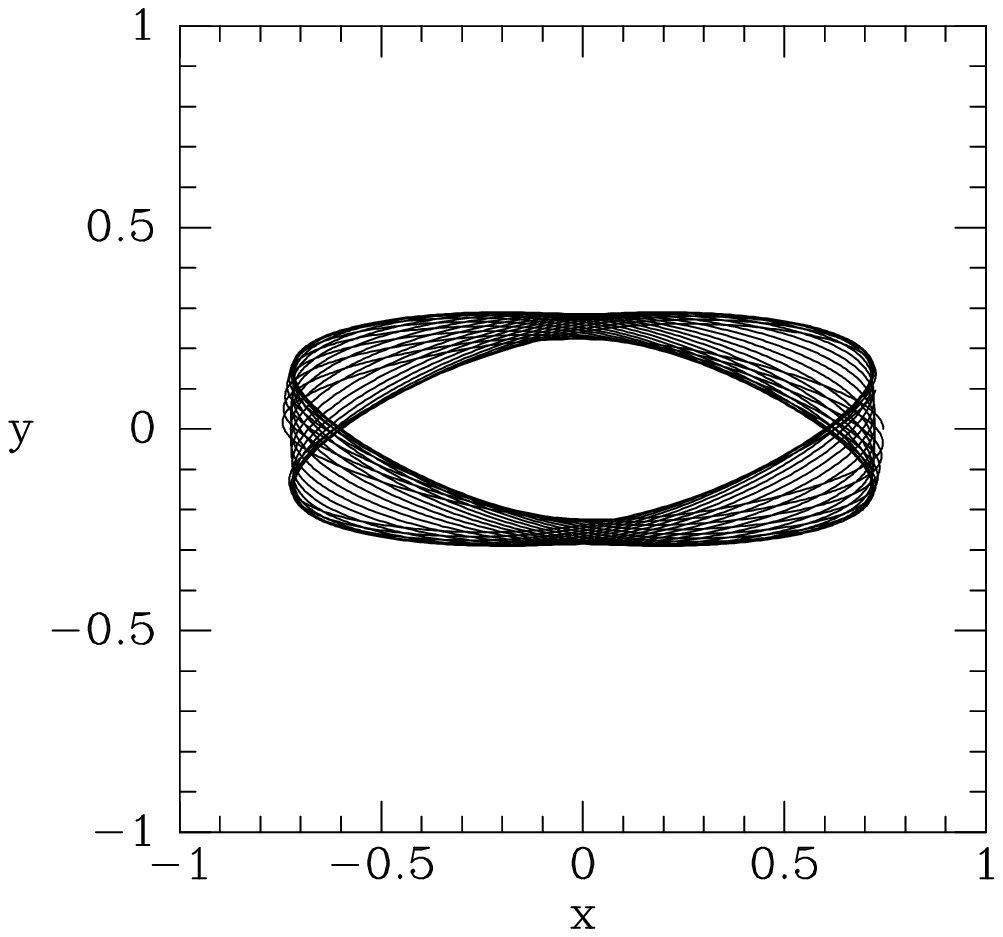}}
\caption{Left panel: closed orbits; $x_1$ orbits black, $x_2$ orbits red.
Centre and right
panels: an orbit that circulates inside the ILR and an orbit trapped by the
ILR  computed using {\tt resTorus\_L}.
The potential has {\tt Bstrength=0.1}.}\label{fig:closed}
\end{figure*}

The left panel of Fig.~\ref{fig:closed} shows the closed orbits in the
potential with ${\tt Bstrength=0.1}$ constructed by integrating the full
equations of motion. The centre and right panels show two orbits constructed
from a common {\tt resTorus\_L}. The orbit in the centre panel circulates
inside the ILR (i.e., in the top panel of Fig.~\ref{fig:ILRHeff} it moves
around the middle red dot). This orbit is clearly librating around an $x_2$
orbit. The orbit in the right panel is trapped at ILR; in the top panel of
Fig.~\ref{fig:ILRHeff} it moves around the red dot on the extreme left. This
orbit can be considered to be librating around an orbit of the $x_1$ family.
A strong bar such as the S15 bar in the cite{PJM11:mass} potential may not
support $x_2$ orbits. This failure evidently arises $x_2$ orbits are orbits
that circulate inside the ILR and all the rungs of a strong bar's ILR ladder
may ladder touch the $J_\phi$ axis, leaving no scope for circulation inside
the ILR.

\begin{figure}
\centerline{\includegraphics[width=.9\hsize]{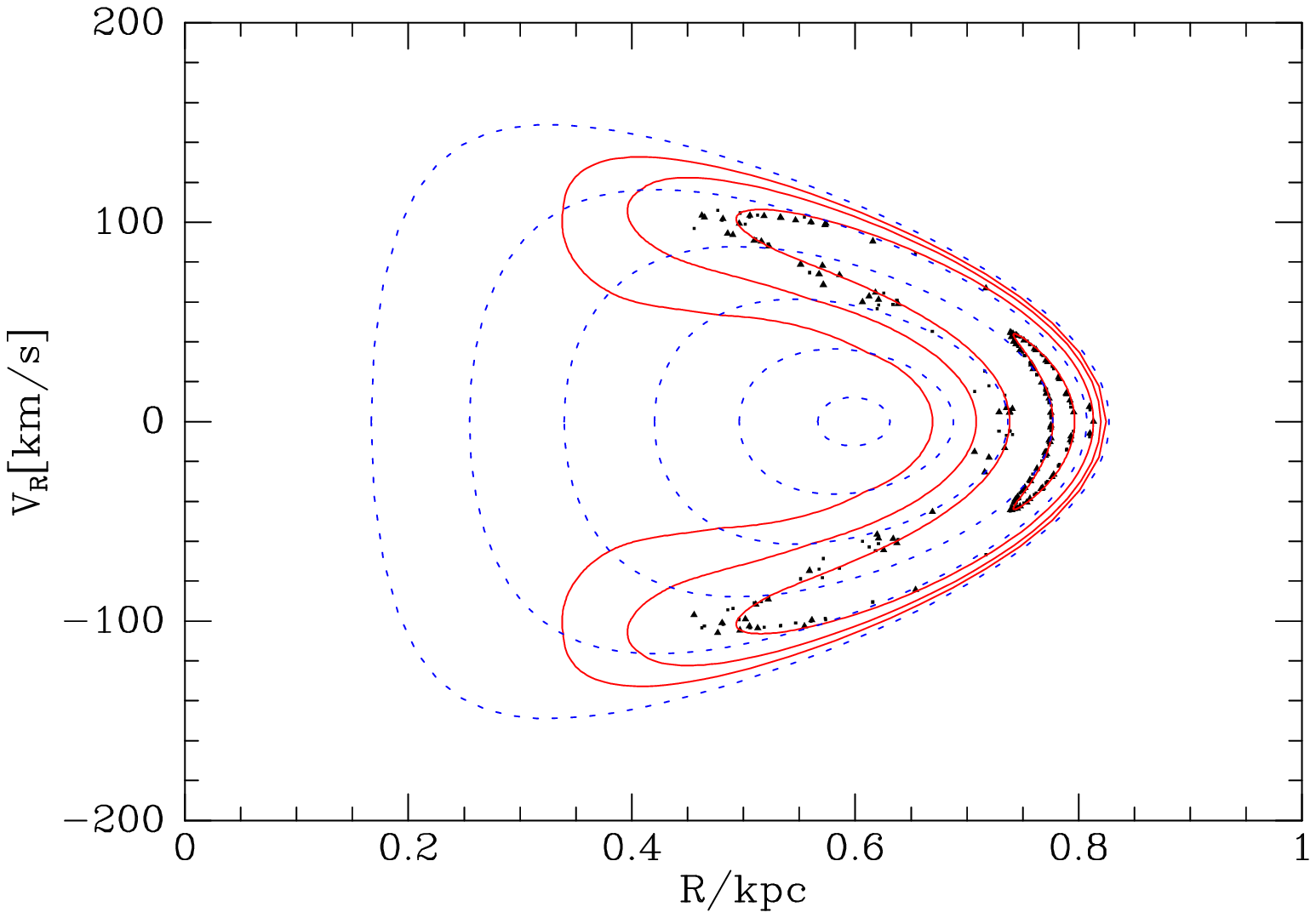}}
\centerline{\includegraphics[width=.9\hsize]{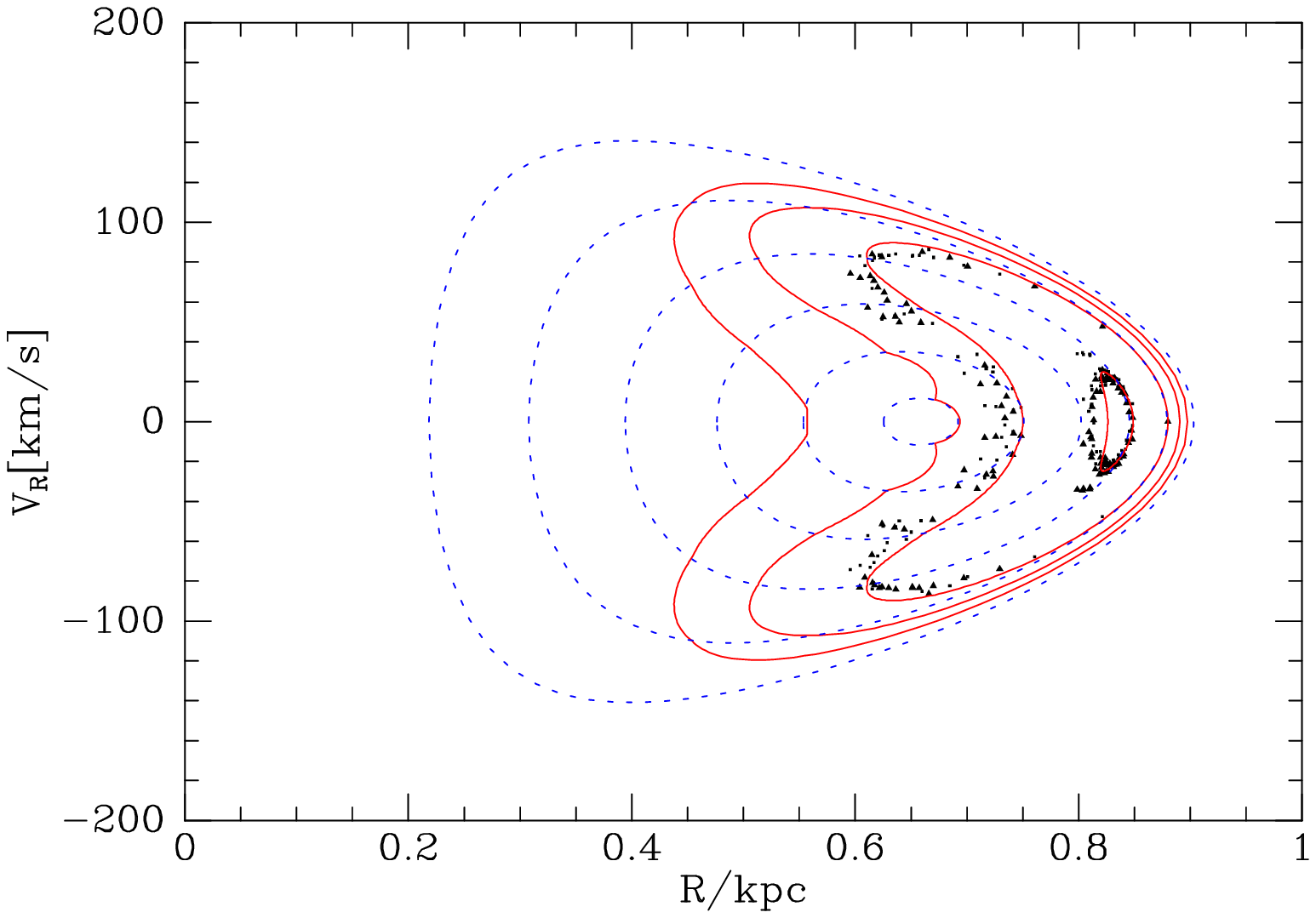}}
\caption{Surfaces of section for orbits trapped at the
ILR. Red curves are
computed from non-axisymmetric  tori constructed using the axisymmetric tori
whose cross sections are plotted in dashed blue curves. Black points show
consequents obtained by numerically integrating the full equations of motion
from initial conditions given by one point on two of the red curves. The
lower panel shows orbits on the lowest rung of the ladder in the bottom panel
of Fig.~\ref{fig:ILR_ladder}, and the upper panel shows orbits on  a higher rung.
}\label{fig:ILRSoS}
\end{figure}

Each panel of Fig.~\ref{fig:ILRSoS} shows sections through tori associated
with a rung of the ladder in the bottom panel of Fig.~\ref{fig:ILR_ladder}. As in
Fig.~\ref{fig:OLRSoS} the blue dashed curves are sections through the
axisymmetric tori that provide the basis for the perturbation theory. The red
curves of each panel are sections through the tori returned by a single {\tt
resTorus\_L} object for five values of the integral $I$. The black dots are
consequents obtained by integrating the full equations of motion for $147\Gyr$
from initial conditions provided by two of these tori. In contrast to
Fig.~\ref{fig:OLRSoS}, the dots do not lie on curves, indicating that the
orbits are stochastic. The degree of stochasticity increases with the
amplitude of libration. The boundaries of the region within which the
consequents of an orbit scatter appear to be bounded by curves belonging to
the family of red curves.

\section{Relation to real-space perturbation theory}\label{sec:Etheory}

Elementary treatments of perturbation of near-circular orbits by a
non-axisymmetric component of the potential \citep[e.g.][\S3.3.3]{GDII} do
not use angle-action coordinates. They proceed by perturbing Newton's
equations of motion to derive the equation of motion for the radial
displacement $\Delta_R$ of an orbiting star that is just the equation of
motion of a driven harmonic oscillator. The particular integral of this
equation corresponds to a closed orbit while the complementary function
describes libration about this orbit. The orientation of the closed orbit
provided by the particular integral switches through $90^{\rm deg}$ across
the resonance, but $\Delta_R$ diverges as the resonance is approached.
Sufficiently far from the resonance these solutions provide useful
approximations to orbits such as those plotted in the right two panels of
Fig.~\ref{fig:closed}. 

These solutions apply to the regime of low $J_r$ in which there is no clear
distinction between orbits that have or have not been trapped. Hence for
angular momenta that are greater than that of the resonant orbit with
$J_r=0$, elementary theory provides an approximation that is valid until its
amplitude $\Delta_R$ becomes a significant fraction of $R$. For angular
momenta smaller than that of the resonant orbit, the approximation is useful
near the part (if any) of the $J_\phi$ axis that lies below and to the left
of the resonance's ladder. The solutions do not describe orbits that lie
within the ladder at values of $J_3'$ below the critical value at which
trapping becomes sharply defined.

\section{Conclusions}\label{sec:conclude}

The algorithm used by \TM\ to compute tori trapped at a Lindblad resonance
has been improved so it can handle trapping at a Lindblad resonance of the
least eccentric orbits. This step was necessary before \TM\ could be used to
investigate the impact of such resonances on the kinematics of the solar
neighbourhood, which is the subject of a companion paper
\citep{Binney:Vspace}. The key change is
to switch from angle-action coordinates, which are polar coordinates for the
slow plane, to Cartesian coordinates for the plane. This switch is only
mandatory for orbits of low eccentricity, but it is expedient to make it for
all orbits.

Orbits that are profoundly affected by a Lindblad resonance occupy an
elongated volume of action space that slopes towards lower angular momentum
$J_\phi$ with increasing radial action, $J_r$. In this zone the vertical
action $J_z$ and a linear combination $J_3'$ of $J_r$ and $J_\phi$ is
conserved to a good approximation, so orbits move along lines of constant
$(J_z,J_3')$. These lines constitute the rungs of a ladder. A new action
$\cJ$ quantifies the magnitude of the excursions an orbit takes from the
ladder's centre line.  

The higher rungs of the ladder have well defined
endpoints at which orbits cease to move to and fro across the ladder. We say
that orbits that are located above or below the ladder circulate rather than
librate.

A number of rungs near the bottom of the ladder reach the $J_\phi$ axis.
Consequently, at the corresponding values of $J_3'$ it is impossible to
circulate inside the resonance. In contrast to rungs higher up the ladder,
the lower rungs do not have well defined upper ends at which libration
suddenly gives way to circulation. Instead as the value of $\cJ$ increases,
there is a continuous evolution of the orbit from libration around an orbit
of a certain eccentricity that differs from an axisymmetric orbit only in
that it does not precess, to motion that is qualitatively the same as that of
an eccentric orbit in an axisymmetric potential.

Our numerical examples have used the potential defined by B18, which is a
combination of a realistic axisymmetric Galaxy model and an analytic
quadrupole. From the likely radius of corotation outwards this potential is
barely differs from that fitted by S15 to the flow of gas through the
Galactic disc. For this potential \TM\ furnishes with ease tori for
corotation and outwards that reproduce with precision the results of direct
integration of the equations of motion.

At $R\la 3\kpc$ the potential's quadrupole is smaller than that of S15, which
is too strong to permit $x_2$ orbits in the axisymmetric potential adopted by
B18.  In the vicinity of the ILR, both bars induce significant chaos in phase
space, and are difficult to use in conjunction with \TM\ because they
require very eccentric tori. So our numerical examples of trapping at the ILR
have been limited to a bar that has a tenth of the strength of the B18 bar.
At this strength, there is limited chaos so one can study the structure of
phase space cleanly.

In a companion paper we use \TM\ to understand better the orbits of stars
that approach the Sun and may be trapped at either corotation or the OLR.  It
is currently unclear what scope \TM\ has to assist in understanding phenomena
deep inside the bar, where trapping by several resonances, especially the
ILR, must be  important. The indications are that orbits in this region are
stochastic, but they are certainly not ergodic. So we have to find means to
quantify them, and \TM, or an upgrade of it,\ is a strong contender for this
job. Given how profoundly orbits inside the bar differ from axisymmetric
orbits, it is tempting to write code that produces triaxial tori directly
rather than distorting axisymmetric images into triaxial tori by perturbation
theory. On the other hand, surprisingly, the signs are that the limitations
encountered here when applying \TM\ inside the bar have less to do with the
perturbation theory than with production of the underlying axisymmetric tori.
In any event, the basic \TM\ code requires further work, in particular to
enable it to interpolate cleanly between tori when a point
transformation is required.

\section*{Acknowledgements}

I thank John Magorrian for an invaluable conversation.
This work has been supported
by the Leverhulme Trust and the UK Science and Technology Facilities Council
under grant number ST/N000919/1.

\bibliographystyle{mn2e} \bibliography{/u/tex/papers/mcmillan/torus/new_refs}

\begin{thebibliography}{}

\bibitem[\protect\citeauthoryear{{Binney}}{{Binney}}{2012}]{JJB12:Stackel}
{Binney} J.,  2012, \mnras, 426, 1324

\bibitem[\protect\citeauthoryear{{Binney}}{{Binney}}{2016}]{Binney2016}
{Binney} J.,  2016, \mnras, 462, 2792

\bibitem[\protect\citeauthoryear{{Binney}}{{Binney}}{2018}]{Binney2018}
{Binney} J.,  2018, \mnras, 474, 2706

\bibitem[\protect\citeauthoryear{{Binney}}{{Binney}}{2020}]{Binney:Vspace}
{Binney} J.,  2020, \mnras, submitted

\bibitem[\protect\citeauthoryear{{Binney} \& {McMillan}}{{Binney} \&
  {McMillan}}{2016}]{JJBPJM16}
{Binney} J.,  {McMillan} P.~J.,  2016, \mnras, 456, 1982

\bibitem[\protect\citeauthoryear{{Binney} \& {Tremaine}}{{Binney} \&
  {Tremaine}}{2008}]{GDII}
{Binney} J.,  {Tremaine} S.,  2008, {Galactic Dynamics: Second Edition}.
Princeton University Press

\bibitem[\protect\citeauthoryear{{Contopoulos} \&
  {Papayannopoulos}}{{Contopoulos} \&
  {Papayannopoulos}}{1980}]{ContopoulosP1980}
{Contopoulos} G.,  {Papayannopoulos} T.,  1980, \aap, 92, 33

\bibitem[\protect\citeauthoryear{{de Zeeuw}}{{de Zeeuw}}{1985}]{deZ85}
{de Zeeuw} T.,  1985, \mnras, 216, 273

\bibitem[\protect\citeauthoryear{{Dehnen}}{{Dehnen}}{2000}]{WD00:OLR}
{Dehnen} W.,  2000, \aj, 119, 800

\bibitem[\protect\citeauthoryear{{Ferri{\`e}re}, {Gillard} \&
  {Jean}}{{Ferri{\`e}re} et~al.}{2007}]{Ferriere2007}
{Ferri{\`e}re} K.,  {Gillard} W.,    {Jean} P.,  2007, \aap, 467, 611

\bibitem[\protect\citeauthoryear{{Kalnajs}}{{Kalnajs}}{1977}]{Kalnajs1977}
{Kalnajs} A.~J.,  1977, \apj, 212, 637

\bibitem[\protect\citeauthoryear{{Lynden-Bell} \& {Kalnajs}}{{Lynden-Bell} \&
  {Kalnajs}}{1972}]{LLKa72}
{Lynden-Bell} D.,  {Kalnajs} A.~J.,  1972, MNRAS, 157, 1

\bibitem[\protect\citeauthoryear{{McMillan}}{{McMillan}}{2011}]{PJM11:mass}
{McMillan} P.~J.,  2011, \mnras, 414, 2446

\bibitem[\protect\citeauthoryear{{Monari}, {Kawata}, {Hunt} \&
  {Famaey}}{{Monari} et~al.}{2017}]{MonariHerc2017}
{Monari} G.,  {Kawata} D.,  {Hunt} J.~A.~S.,    {Famaey} B.,  2017, \mnras,
  466, L113

\bibitem[\protect\citeauthoryear{{P{\'e}rez-Villegas}, {Portail}, {Wegg} \&
  {Gerhard}}{{P{\'e}rez-Villegas} et~al.}{2017}]{Perez2017}
{P{\'e}rez-Villegas} A.,  {Portail} M.,  {Wegg} C.,    {Gerhard} O.,  2017,
  ArXiv e-prints

\bibitem[\protect\citeauthoryear{{Sanders} \& {Binney}}{{Sanders} \&
  {Binney}}{2014}]{SaJJB14}
{Sanders} J.~L.,  {Binney} J.,  2014, \mnras, 441, 3284

\bibitem[\protect\citeauthoryear{{Sanders} \& {Binney}}{{Sanders} \&
  {Binney}}{2015}]{SaJJB15:Triaxial}
{Sanders} J.~L.,  {Binney} J.,  2015, \mnras, 447, 2479

\bibitem[\protect\citeauthoryear{{Sormani}, {Binney} \& {Magorrian}}{{Sormani}
  et~al.}{2015}]{SormaniIII}
{Sormani} M.~C.,  {Binney} J.,    {Magorrian} J.,  2015, \mnras, 454, 1818

\bibitem[\protect\citeauthoryear{{Weinberg}}{{Weinberg}}{2001}]{Weinberg2001}
{Weinberg} M.~D.,  2001, \mnras, 328, 311

\end{thebibliography}

\end{document}